\documentclass[twocolumn,showpacs,preprintnumbers,amsmath,amssymb,floatfix]{revtex4}
\usepackage{graphicx}
\usepackage{dcolumn}
\usepackage{bm}

\newcommand\HR{\rule{0em}{0pt}}
\newcommand{\be}{\begin{equation}}
\newcommand{\ee}{\end{equation}}
\newcommand{\ba}{\begin{eqnarray}}
\newcommand{\ea}{\end{eqnarray}}
\newcommand{\ban}{\begin{eqnarray*}}
\newcommand{\ean}{\end{eqnarray*}}

\newcommand{\hu}{\hat{U}}
\newcommand{\cu}{{\cal U}}
\newcommand{\hv}{\hat{V}}

\newcommand{\tta}{\tilde{\tau}}
\newcommand{\uu}[1]{{u}_{#1}}
\newcommand{\uuu}[1]{{u}^{(2)}_{#1}}
\newcommand{\pa}{\partial}

\newcommand{\su}{\sum_{n=0}^{\infty}}

\newcommand{\n}[1]{\label{#1}}

\newcommand{\non}{\nonumber}

\newcommand{\eq}[1]{(\ref{#1})}

\newcommand{\tpsi}{\tilde{\psi}}
\newcommand{\ind}[1]{\mbox{\tiny #1}}

\newcommand{\hh}{\, ,\hspace{0.5cm}}
\newcommand{\hhh}{\, ,\hspace{0.2cm}}

\begin{document}
 
\title{Interior of Distorted Black Holes}
\author{Valeri P. Frolov}
\email{frolov@phys.ualberta.ca}
\author{Andrey A. Shoom}
\email{ashoom@phys.ualberta.ca}
\affiliation{Theoretical Physics Institute, University of Alberta, 
Edmonton, AB, Canada,  T6G 2G7}
\date{\today}
\begin{abstract}
We study the interior of distorted static axisymmetric black holes. We obtain a general interior solution and study its asymptotics both near the horizon and singularity. As a special
example, we apply the obtained results to the case of the so-called `caged' black holes.    
\end{abstract}

\pacs{Valid PACS appear here \hfill  
Alberta-Thy-18-06}
\pacs{04.20.Dw, 04.20.Cv, 04.70.Bw}
\maketitle
\section{\label{sec:level1}INTRODUCTION}

The uniqueness theorem proved by Israel \cite{Israel} tells us that
the only static vacuum black hole solution of the Einstein equations
in an asymptotically flat spacetime is the Schwarzschild one.  In the application to a real astrophysical problem this solution, even in the absence of rotation, is highly idealized. For
example, the presence of matter, e.g. in the form of accretion disk,
distorts the metric.  If a static distribution of matter is
localized outside the black hole horizon, the spacetime in the vicinity of the
horizon remains vacuum. We call such a solution a {\em distorted black
hole}. The metric near the horizon  of a general (not necessary axisymmetric)
static distorted black hole was studied in \cite{FrSa}. 

If the distribution of matter outside a black hole is axisymmetric,
the metric of the distorted black hole allows detailed description. The reason is that the vacuum metric
outside the matter is the Weyl solution. This metric contains two
functions of two variables. One of this function, which has the
meaning of the gravitational potential, obeys the linear Laplace
equation in a flat 3D space, while the other can be obtained from it
by a simple integration. Axially symmetric distorted black holes were
studied in several publications (see e.g.
\cite{IsKh,Isra:73,MySz,Pet,Geroch,Chandrabook,FaKr:01}).    Such
axially symmetric distorted black holes arise naturally in the models
were one of the (large) spatial dimensions is compactified. For the
general discussion of such solutions in higher dimensions see, e.g.
\cite{Myers:87,HaOb:02}. In 4D such caged black hole solution
is again the Weyl metric. The properties of 4D caged black holes were
studied in \cite{BoPe:90,FrFr}. 

In the previous studies of distorted black holes the attention has mainly 
been focused on the properties of the black hole exterior.
But any distribution of matter in the black hole exterior region distorts the
metric not only outside the black hole, but  also in its interior. The
purpose of this paper is to study this effect. Namely, we consider
the interior of an axially symmetric distorted black hole. In
particular, we study the structure of the spacetime in the vicinity
of the black hole singularity.

The paper is organized as follows. In Section~2 we collect the
equations for the vacuum axisymmetric distorted black hole in the
exterior and interior regions.  In Section~3 we obtain solution
for the interior of distorted black hole  and discuss its
properties. An asymptotic form of this solution near the black hole horizon and singularity is obtained in Section~4 and Section~5, respectively.
Special examples of exact interior solutions and their properties are
considered in Section~6. In Section~7 we consider properties of the
interior and singularity of 4D caged black hole. Section~8
contains summary and discussions of the results obtained .  
Additional technical details and calculations used in the main part
of the paper are collected in Appendices. In this paper we use the 
units where $G=c=1$, and the sign conventions adopted in \cite{MTW}.

\section{Metric of a distorted black hole}

\begin{figure}[htb]
\begin{center} 
\includegraphics[height=4.5cm,width=4.5cm]{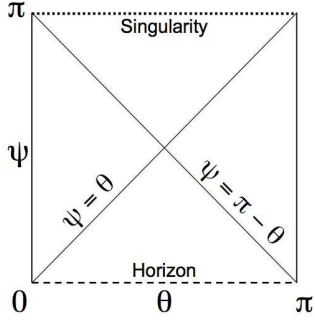} 
\caption{The Carter-Penrose diagram for the $(\psi$-$\theta$)-sector of the Schwarzschild black hole interior.} \label{f1} 
\end{center}
\end{figure}

In the absence of distortion a static vacuum black hole is described
by the Schwarzschild metric 
\be
ds^2=-(1-2m/r)dt^2+\frac{dr^2}{1-2m/r}+r^2 d\omega^2\, ,
\ee
where $m$ is the black hole mass, and $d\omega^2=d\theta^{2}+\sin^{2}\theta
d\phi^{2}$ is the metric on a unit round sphere.
In what follows, we shall use  two other forms of this metric 
\begin{eqnarray}
ds^{2}&=&-\frac{\cosh\tpsi -1}{\cosh\tpsi +1}dt^{2}+m^{2}
(\cosh\tpsi+1)^{2} (d\tpsi^{2}+d\omega^2) \, ,\label{ex}\\
ds^{2}&=&\frac{1-\cos\psi}{1+\cos\psi} dt^{2}+m^{2}
(1+\cos\psi)^{2}(-d\psi^{2}+d\omega^2)   \, .\label{in}\hspace{0.4cm}
\end{eqnarray}  
The metric \eq{ex} is valid outside the Schwarzschild black hole horizon, $r>2m$, and
$r=m(\cosh\tpsi+1)$, where $\tpsi >0$, while the metric \eq{in} covers the Schwarzschild 
black hole interior $0<r<2m$, and $r=m(1+\cos\psi)$, where $0<\psi<\pi$. These two 
metrics are connected by the analytical continuation
\begin{equation}
\tpsi\to i\psi\label{eq1a} .
\end{equation} 
The Carter-Penrose diagram for the interior metric in the
coordinates $(\psi,\theta)$ is shown on Figure \ref{f1}. 
The lines $\psi\pm\theta=const$ are null rays propagating within the 2D
section $(t,\phi)=const$.
It should be stressed that this diagram is different from the usual
Carter-Penrose diagram for the radial sector $(t,r)$ of the Schwarzschild black hole.

Following  \cite{Geroch,Chandrabook} we present the metric for the
vacuum axisymmetric static distorted black hole in the form 
\begin{eqnarray}
ds^{2}&=&-\frac{\cosh\tpsi -1}{\cosh\tpsi +1}e^{2\hat{U}}dt^{2}+m^{2}(\cosh\tpsi+1)^{2}e^{-2\hat{U}}\nonumber\\
&\times& \left[ e^{2\hat{V}}\left(d\theta^{2}+d\tpsi^{2}\right)+\sin^{2}
\theta d\phi^{2}\right] \, ,\label{eq2e}
\end{eqnarray}
\begin{eqnarray}
ds^{2}&=&\frac{1-\cos\psi}{1+\cos\psi}e^{2\hat{U}}dt^{2}+m^{2}
(1+\cos\psi)^{2}e^{-2\hat{U}}\nonumber\\
&&\times \left[ e^{2\hat{V}}\left(d\theta^{2}-d\psi^{2}\right)+\sin^{2}
\theta d\phi^{2}\right] \, .\label{eq2}
\end{eqnarray}
The metric \eq{eq2e}, where $\hu$ and $\hv$ are functions of
$(\tpsi,\theta)$, is valid in the black hole exterior. The metric
\eq{eq2}, where $\hu$ and $\hv$ are functions of $(\psi,\theta)$,
describes the interior of the distorted black hole. The black hole
horizon $H$ is defined by the equation
\be
\tpsi=\psi=0\hh 0\le \theta\le \pi\, .
\ee
For a regular black hole the functions $\hat{U}$ and  $\hat{V}$
must be smooth and finite at the horizon. In particular, these
functions, which define the geometry of the 2D horizon
surface, must be continuos across the horizon. The metrics \eq{eq2e}
and \eq{eq2} are regular along the axis of symmetry (no conical
singularities), provided
\be\n{V0}
\hat{V}|_{\theta=0}=\hat{V}|_{\theta=\pi}=0\, .
\ee
Denote
\ba
D_{\psi}\hu&=&\hat{U}_{,\psi\psi}+\cot\psi \,\hat{U}_{,\psi}\,
,\n{Dpsi}\\ 
D_{\theta}\hu&=&\hat{U}_{,\theta \theta}+
\cot\theta\,\hat{U}_{,\theta}\, .
\n{Dtheta}
\ea
Then, the vacuum Einstein equations for the metric \eq{eq2} reduce to the
following relations
\ba
D_{\psi}\hu&=&D_{\theta}\hu\, ,\n{DD}\\
\hat{V}_{,\theta}&=&F^{(\theta)}(\psi,\theta)\\
\hat{V}_{,\psi}&=&F^{(\psi)}(\psi,\theta)
\end{eqnarray}
Here
\ba
F^{(\theta)}(\psi,\theta)&=&N
\left[\sin^{2}\psi\cos\theta\left(\hat{U}^{2}_{,\theta}+\hat{U}^{2}_{,\psi}\right)
\right.\nonumber\\
&& - \ 2\sin\psi\cos\psi\sin\theta\, \hat{U}_{\theta}\, \hat{U}_{\psi}\label{eq3a}\\
&& + \ 2\sin\psi\cos\theta
\left.\hat{U}_{,\psi}-2\cos\psi\sin\theta\, \hat{U}_{,\theta}\right],\nonumber\\
F^{(\psi)}(\psi,\theta)&=&N
\left[2\sin^{2}\psi\cos\theta \, 
\hat{U}_{,\theta}\hat{U}_{,\psi}\right.\ \ \ \ \ \ \nonumber\\
&& - \ \sin\psi\cos\psi\sin\theta\left(\hat{U}^{2}_{,\theta}
+ \hat{U}^{2}_{,\psi}\right) \label{eq3b} \\
&& \left. + \  2\sin\psi\cos\theta\hat{U}_{,\theta}
-2\cos\psi\sin\theta\, \hat{U}_{,\psi}
\right]\, ,\label{eq3c}\nonumber\\
N&=&\sin\theta (\sin^{2}\psi-\sin^{2}\theta)^{-1} \, .
\end{eqnarray}
The functions $F^{(\theta)}$ and $F^{(\psi)}$ obey the relation
\be
F^{(\theta)}_{,\psi}=F^{(\psi)}_{,\theta}\, .
\ee
Thus, after solving equation \eq{DD}, one can obtain $\hv$ by
simple integration
\be\n{intv}
\hv(\psi,\theta)=\int_{(\psi_0,\theta_0)}^{(\psi,\theta)}
[F^{(\psi)}d\psi+F^{(\theta)} d\theta]\, ,
\ee
where the integral is taken along any path connecting
$(\psi_0,\theta_0)$ and $(\psi,\theta)$.
 
The factor $N$ is singular along the lines $\psi=\theta$
and $\psi=\pi-\theta$. Nevertheless, as we shall demonstrate in the next
Section, the solutions for $\hv$ for the distorted black holes which
are regular at the horizon remain smooth and regular along these lines.

Let us focus on equation \eq{DD}. Since
$D_{\psi}\hu=D_{-\psi}\hu$, a solution to this equation can be presented as
a sum of two solutions, one being odd and the other being
even function of $\psi$. Because of the presence of factor
$\cot\psi$ in  $D_{\psi}$, this operator is singular at $\psi=0$, hence,
the regular at the horizon solution must be an even function of
$\psi$. This solution remains real after the Wick's rotation
\eq{eq1a}.

Studying the exterior solution
\be\n{uex}
D_{\tilde{\psi}}\hu + D_{\theta}\hu=0\, 
\ee 
Geroch and Hartle \cite{Geroch}
demonstrated that if $\hat{U}$ is a regular smooth function in any small open neighborhood 
of $H$ (including $H$ itself) which takes the same values, $u_0$, on the both ends of the
segment $H$, 
\begin{equation}
\hu(0,0)=\hu(0,\pi)=u_0, \label{eq4b}
\end{equation}
then the solution is regular at the horizon and
describes a distorted black hole. 

The surface gravity of the distorted
black hole is
\be
\kappa_0=\frac{e^{2u_0}}{4m}\, .
\ee
If the distortion source obeys the strong energy condition $u_0$ has to be non-positive \cite{Geroch}.

Let us emphasize that equation \eq{uex} for the `gravitational potential'
$\hu$ in the  exterior region is elliptic,
while the interior equation \eq{DD} is of the hyperbolic type. This
is in accordance with the general property of black holes. Namely,
direction to the singularity in the inner region is the direction to
the future, and the evolution of the metric in this region obeys
dynamical equations. After solving the equations
in the exterior region we obtain boundary conditions at the horizon
for the inner dynamical equations. In our particular case the
solutions of the exterior and interior problems are connected by the
analytical continuation \eq{eq1a}. 

It is convenient to consider dimensionless form of
the metric $dS^2$, connected to the metric $ds^2$ as follows
\be
ds^2=4m^2 e^{-2u_0} dS^2\, .
\ee 
Introducing the quantities
\be\n{sc}
T=\kappa_0 t\hh {\cal U}=\hat{U}-u_0\, ,
\ee
one can write the metric $dS^2$ in the form
\begin{eqnarray}
dS^{2}&=&4 \frac{1-\cos\psi}{1+\cos\psi}e^{2\cu} dT^{2}+\frac{1}{4}
(1+\cos\psi)^{2}e^{-2\cu}\nonumber\\
&&\times \left[ e^{2\hat{V}}\left(d\theta^{2}-d\psi^{2}\right)+\sin^{2}
\theta d\phi^{2}\right] \, .\label{mm}
\end{eqnarray}
In what follows, we shall study the metric \eq{mm} and its properties.
In order to obtain the corresponding characteristics of the `physical'
solution \eq{eq2} it is sufficient to use the scaling transformations
\eq{sc}.

\section{Interior solution}

\subsection{Gravitational potential in the inner region}

Our goal is to study the interior of a distorted black hole. 
To find the metric inside the distorted black hole we start with equation (\ref{DD}). This equation allows a separation of variables
$\hu=R(\psi) S(\theta)$ 
\ba
S_{,\theta \theta}+ \cot\theta\,S_{,\theta}+\lambda S=0\, ,\n{S}\\
R_{,\psi\psi}+\cot\psi \,R_{,\psi} +\lambda R=0.\n{R}
\ea
Since the polar points $\theta=0,\pi$ are regular, the functions $S$
must be finite at these points. The solutions of this eigenvalue
problem are 
\be
S=P_n(\cos\theta)\hh\lambda=n(n+1)\hh n=0,1,\ldots\, .
\ee
Expanding $\hu$ over the complete set of Legendre polynomials of the
first kind, $P_n(\cos\theta)$, one has
\be
\hu(\psi,\theta)=\sum_{n\geq 0}R_n(\psi)P_n(\cos\theta)\, .
\ee
Since at the horizon surface $\hu$ is finite and regular, one must
omit infinitely growing at $\psi=0$ solutions of \eq{R}. Thus, we obtain
the following solution for $\hu$
\begin{equation}
\hat{U}(\psi,\theta)=\sum_{n\geq
0}a_{n}P_{n}(\cos\psi)P_{n}(\cos\theta),
\label{eq6}
\end{equation}
where $a_{n}$ are  the coefficients called multipole moments.  For a
given value of $\hu$ on the horizon these coefficients are
\be\n{an}
a_n=(n+1/2)\int_{0}^{\pi}d\theta \sin\theta\, \hu(0,\theta) P_n(\cos\theta)\, .
\ee
The condition
\eq{eq4b} implies
\be
\sum_{k\geq 0}a_{2k+1}=0\hh \sum_{k\geq 0}a_{2k}=u_0\, . \label{eq7b}
\ee

Since the Legendre polynomials have the symmetry property
\be
P_n(-z)=(-1)^n P_n(z)\, ,
\ee
the function $\hu$ is invariant under the transformation
\be\n{sym}
I: (\psi,\theta)\to (\pi-\psi,\pi-\theta)\, ,
\ee
that is
\be\n{symU}
\hu(\pi-\psi,\pi-\theta)=\hu(\psi,\theta)\, .
\ee
This relation implies, in particular, that the value of $\hu$
at the singularity $\psi=\pi$ is determined by its values on
the horizon $\psi=0$
\be
\hu(\pi,\pi-\theta)=\hu(0,\theta)\, .
\ee
In other words, there exists an interesting duality between the
horizon and singularity. It should be emphasized that the functions
$F^{(\theta)}$ and $F^{(\psi)}$, (see \eq{eq3a} - \eq{eq3b}), contain both,
symmetric and antisymmetric parts with respect to the reflection
\eq{sym}. This means that the function $\hv$ does not possesses the
symmetry \eq{symU}. Nevertheless, since the function $\hu$ and
boundary conditions \eq{V0} determine $\hv$ uniquely, the relation
\eq{symU} simplifies greatly the study of the spacetime structure near
the singularity. We return to this point in Section~V.

\begin{figure}[htb]
\begin{center} 
\includegraphics[height=4.5cm,width=4.5cm]{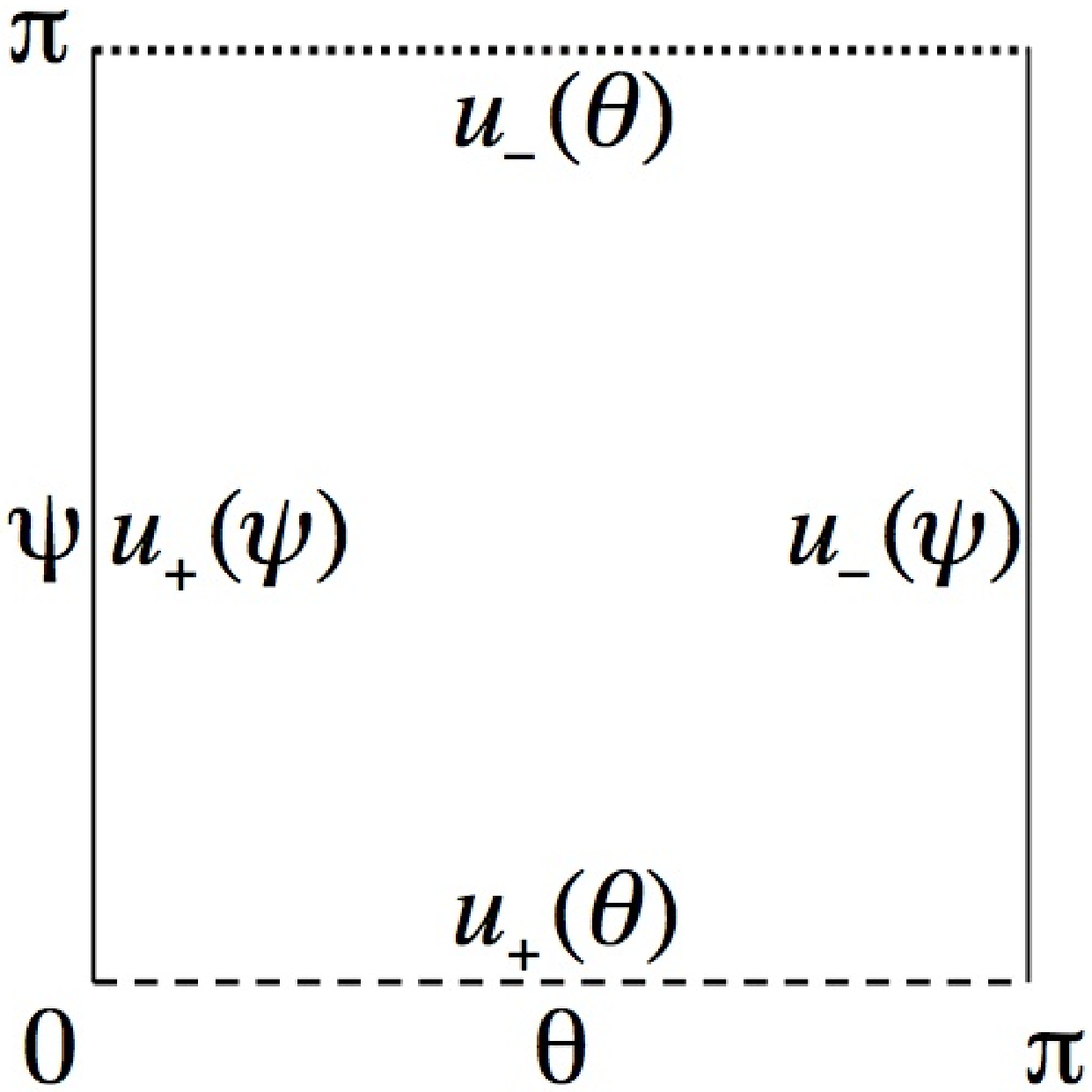} 
\caption{The boundary value of $\cu$ for the  interior of a distorted
black hole. The dashed line,  $\psi=0$, is the horizon, and the dotted
line, $\psi=\pi$, is the singularity.} \label{f2} 
\end{center} 
\end{figure}

\subsection{Boundary values of $\hu$ and $\hv$}

Let us denote
\be
\uu{\pm}(\theta)=\sum_{n\geq 0}(\pm 1)^{n}a_{n}P_{n}(\cos \theta)-u_0.\n{u}
\ee
It is easy to check that
\ba
\uu{\pm}(\theta)&=&\uu{\mp}(\pi-\theta)\, ,\label{f2a}\\
\uu{\pm}(0)&=&\uu{\pm}(\pi)=0\, .
\ea
One has (see Figure \ref{f2})
\ba
\cu(0,\theta)&=&\uu{+}(\theta)\hhh
\cu(\pi,\theta)=\uu{-}(\theta),\label{eqf2b}\\
\cu(\psi,0)&=&\uu{+}(\psi)\hhh
\cu(\psi,\pi)=\uu{-}(\psi).\label{eqf2c}
\ea
We call these data the {\em boundary value of $\cu$}. 
The regularity of $\uu{\pm}$ follows from the regularity of the
distortion $\hu$ on the horizon and the condition (\ref{f2a}). 

\begin{figure}[htb]
\begin{center} 
\includegraphics[height=4.5cm,width=4.5cm]{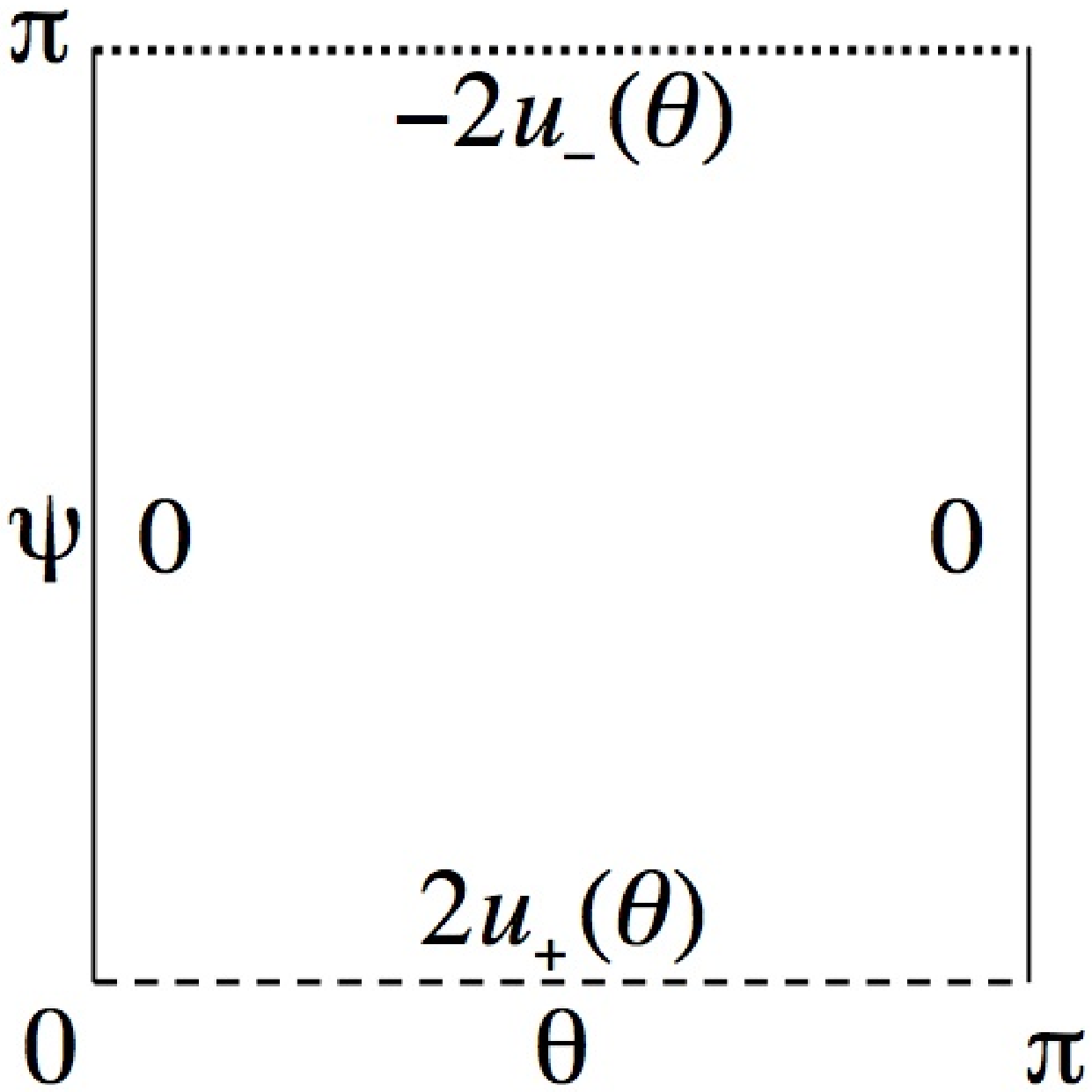} 
\caption{The boundary values of $\hv$ for the  interior of a distorted
black hole. The dashed line,  $\psi=0$, is the horizon, and the dotted
line, $\psi=\pi$, is the singularity.} \label{f3} 
\end{center} 
\end{figure}

The boundary values of $\hv$ can be obtained from \eq{intv}. 
Namely, integrating $F^{(\theta)}$ along the lines $\psi=0$, $\theta_0=0$, and
$\psi=\pi$,  $\theta_0=0$ one obtains (see Appendix~B)
\be
\hv(0,\theta)=2\uu{+}(\theta),\hh \hv(\pi,\theta)=-2\uu{-}(\theta)\, .
\label{eqf2d0}
\ee
The boundary values of $\hv$ are summarized on Figure \ref{f3}.

\subsection{Proper time of free fall to the singularity along the
symmetry axis}

\begin{figure}[htb]
\begin{center}
\HR
\ba
&\includegraphics[height=4.22cm,width=4cm]{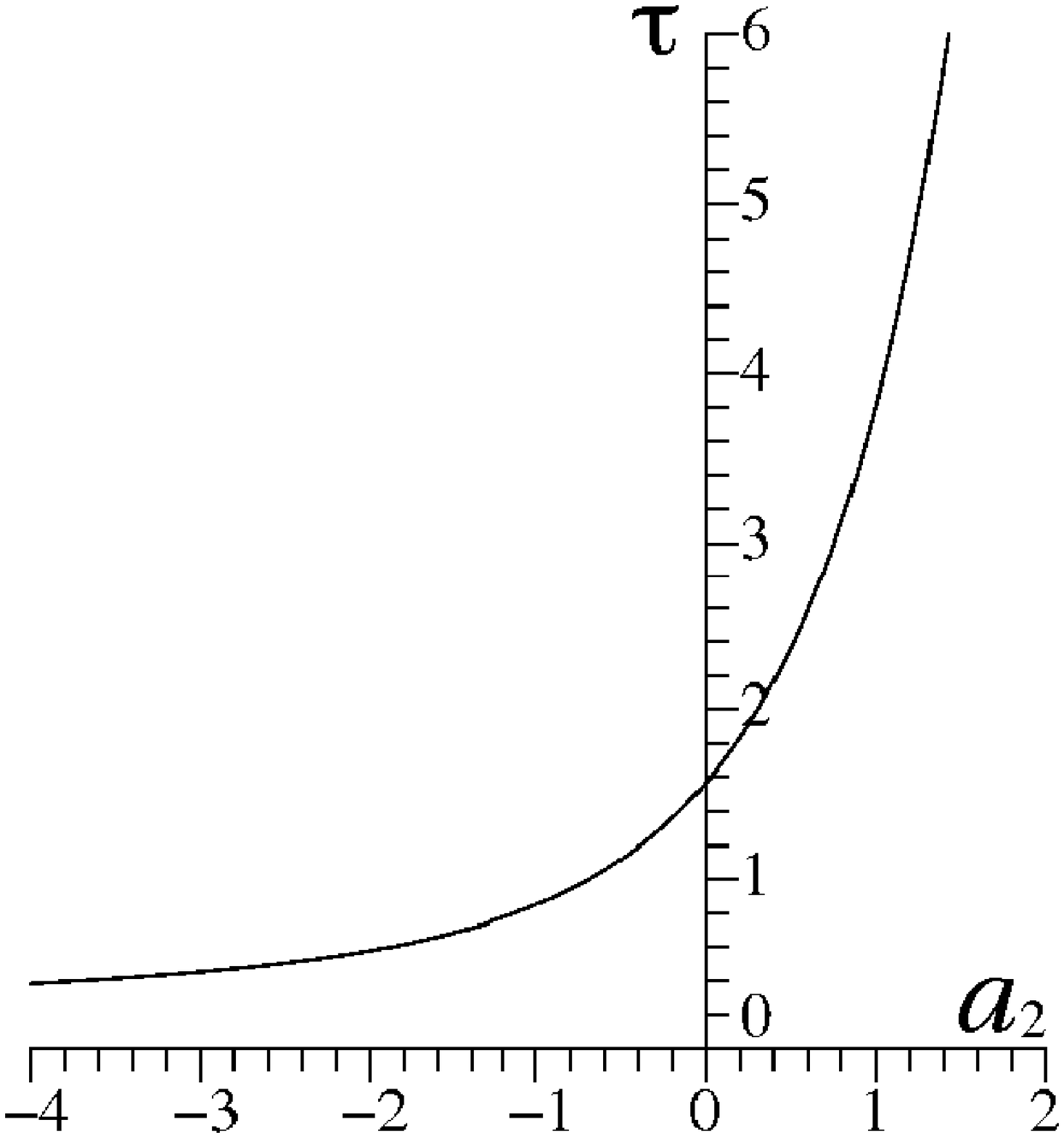}
&\includegraphics[height=4.216cm,width=4cm]{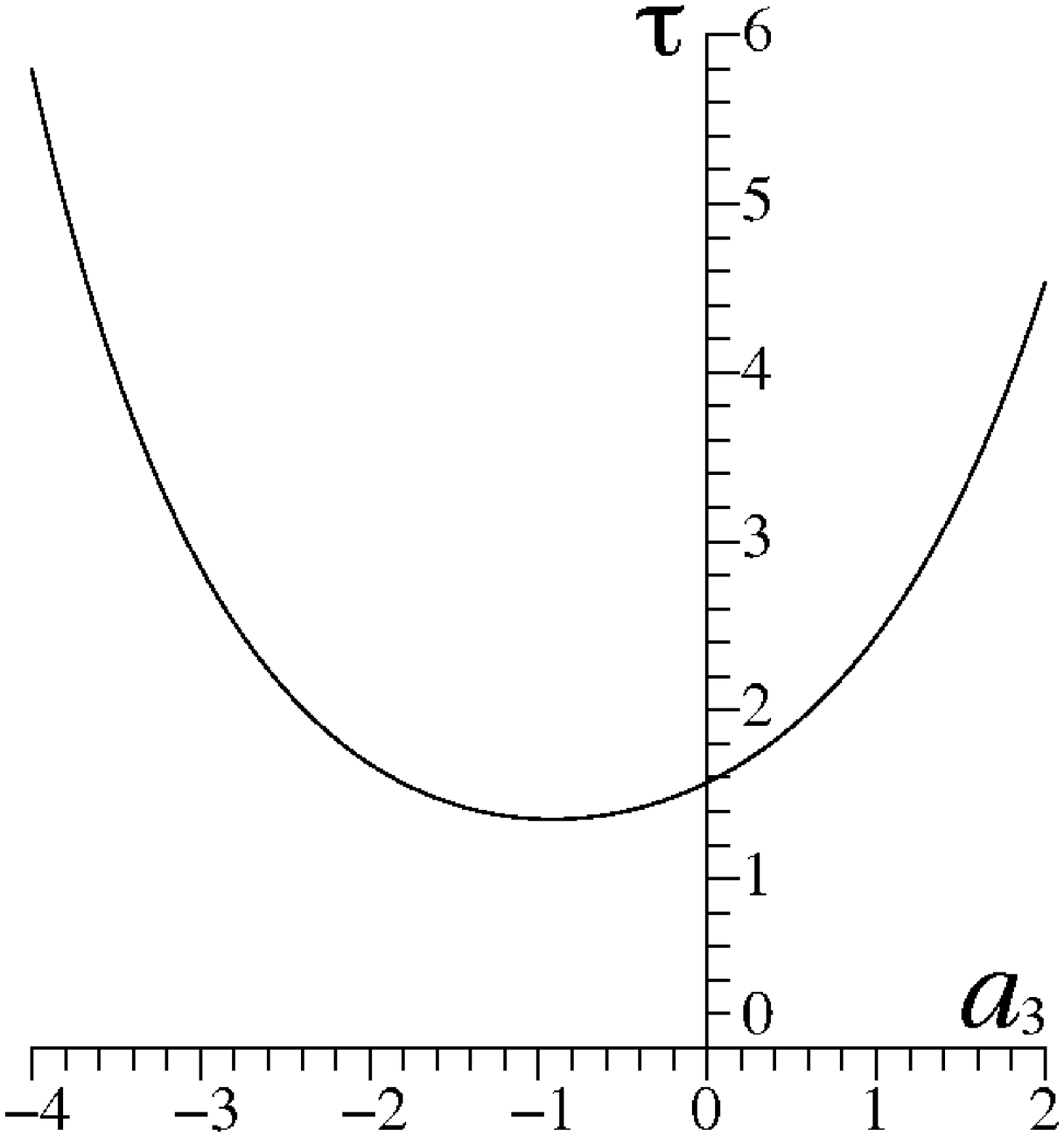}
\nonumber\\
& {\bf a} &\hspace{1.9cm}{\bf b}\nonumber
\ea
\vspace{-0.5cm}
\caption{({\bf a}): Quadrupole distortion: The proper time, $\tau$, as a
function of the quadrupole moment $a_2$. ({\bf b}) Octupole distortion: The proper time, $\tau_+$, as a function of the octupole moment $a_3$. A plot for $\tau_-$ can be obtained from this one by the reflection $a_3\to -a_3$.}
\label{f4} 
\end{center} 
\end{figure}

Suppose a function $u_+(\theta)$ for a distorted black hole is given.
Then the boundary data for $\cu$ and $\hv$ described above allow one
to calculate, for example, a proper time of a free fall  of a test
particle from the horizon to the singularity along the symmetry axis.
Consider a particle with zero energy and angular momentum which is
moving along $T=const$ and $\phi=0$. The proper time $\tau$ of its
free fall from the horizon to the singularity along the symmetry axis  calculated for the metric \eq{mm}
is given by
\be\n{time}
\tau_{\pm}=\frac{1}{2}\int_{0}^{\pi} d\psi (1+\cos\psi)e^{-u_{\pm}(\psi)}\, . 
\ee
The $\pm$ signs are for $\theta=0$ and $\theta=\pi$ axes,
respectively. To illustrate how the distortion of the black hole
affects this time we consider two simple cases. As the first example, we
consider the quadrupole distortion when only $a_0$ and $a_2$ do not vanish.
Taking into account that $a_0=u_0-a_2$ (see \eq{eq7b}) one has
\be\n{uev}
u_{\pm}=-\frac{3}{2}a_2\sin^2\psi\, .
\ee
The integral \eq{time} can be calculated exactly
\be
\tau_{\pm}=\frac{\pi}{2}e^{3a_2/4}I_0(3a_2/4)\, ,
\ee
where $I_0(z)$ is the modified Bessel function. A plot of $\tau$ as a
function of the quadrupole moment $a_2$ is shown on Figure \ref{f4} ({\bf a}).

As a second example we consider the octupole distortion when only
$a_1$ and $a_3$ do not vanish. Because of \eq{eq7b} one has $a_1=-a_3$, and
\be\n{uod}
u_{\pm}=\mp \frac{5}{2}a_3\,\cos\psi\, \sin^2\psi\, .
\ee
To obtain $\tau_{\pm}$ we used numerical integration. Plot of $\tau_+$ as a function of the octupole moment $a_3$ is shown
on Figure \ref{f4} ({\bf b}). The minimal value of $\tau_+$ corresponds to
$a_3\approx -0.9166$. The similar plot for $\tau_-$ can be obtained by
the reflection $a_3\to -a_3$.

\section{Near horizon geometry} 

\subsection{Shape of a distorted horizon}

The form of the horizon surface for the metric \eq{mm} is determined by the following line element (see also \cite{Geroch})
\ba
d\sigma^2_+&=& e^{2\uu{+}} \frac{dz^2}{1-z^2}+
e^{-2\uu{+}}(1-z^2)\, d\phi^2 \non\\
&=& e^{2\uu{+}} d\theta^2+
e^{-2\uu{+}}\sin^2\theta\, d\phi^2\, .
 \label{eq111}\n{dsp}
\ea
This metric is obtained from \eq{mm} as the limit $\psi\to 0$ of the
metric on  2D section $\psi=const$ of $T=const$. The
horizon area is  ${\cal A}=4\pi$.

The Gaussian curvature of the metric $d\sigma^2_+$ is $K_+=R/2$, where
$R$ is the Ricci scalar curvature. It is given by the following
expression
\ba
K_+ &=&e^{-2\uu{+}}\left[1+ (1-z^2)\,[\uu{+}'' -2(\uu{+}')^2]\,-4z \uu{+}' \right],\label{eq19c}\nonumber\\
&=&e^{-2u_{+}}\left(1+u_{+,\theta \theta}+3\cot\theta u_{+,\theta}-2u^{2}_{+,\theta}\right)\, ,
\ea
where the prime denotes a derivative with respect to $z=\cos\theta$.

As the special examples, we consider the quadrupole and octupole distortions with functions $u_+$ given by  \eq{uev} and \eq{uod} respectively.
The Gaussian curvature for these distortions is
\ba
{}^{(2)}K_+&=&e^{3a_2\sin^2\theta}\left[1+3a_2(1-5\cos^2\theta)\right.\nonumber\\
&-&\left. 18a_2^2\cos^2\theta\sin^2\theta\right]\, ,\\
{}^{(3)}K_+&=&\frac{1}{2}e^{5a_3\cos\theta\, \sin^2\theta}\left[2-10a_3\cos\theta(9\cos^2\theta-5)\right.\nonumber\\
&-&\left. 25a_3^2\sin^2\theta(1-3\cos^2\theta)^2\right]\, .
\ea
For the quadrupole distortion the Gaussian curvature becomes negative
at both of the poles, $\theta=0$ and $\theta=\pi$, for $a_2>1/12$. Similarly, for the
octupole distortion, the Gaussian curvature becomes negative at one of
the poles for $|a_3|>1/20$. It means that for these values of the
parameters the horizon surface of the distorted black hole cannot be
isometrically embedded in a flat 3D space (see e.g. \cite{FR} and
references therein). For $a_2\le 1/12$ (in the quadrupole case) and
$|a_3|\le 1/20$ (in the octupole case) isometric embeddings are possible.

\begin{figure}[htb]
\begin{center}
\HR
\ba
&\includegraphics[height=4.29cm,width=4cm]{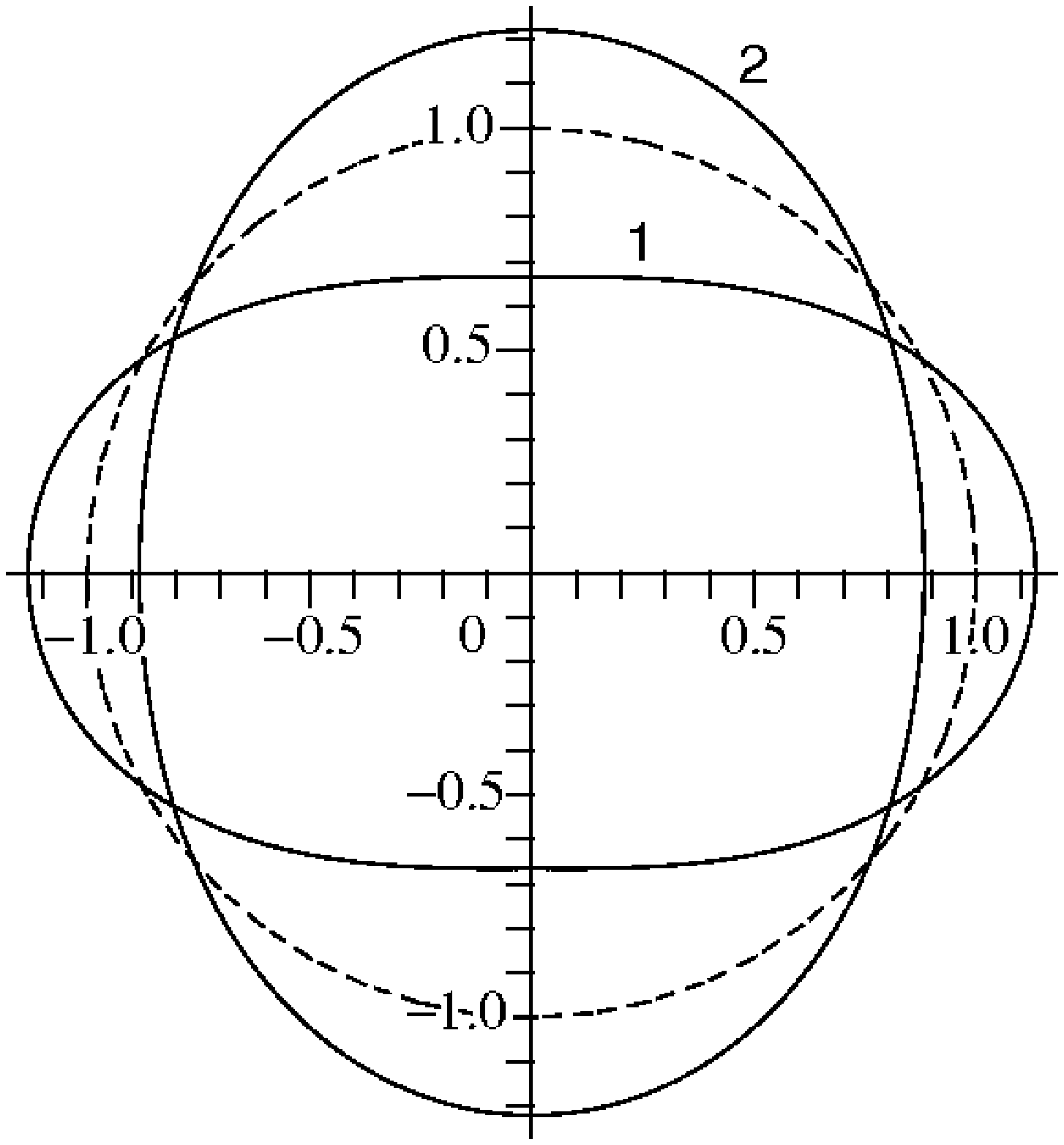}
&\includegraphics[height=4.26cm,width=4cm]{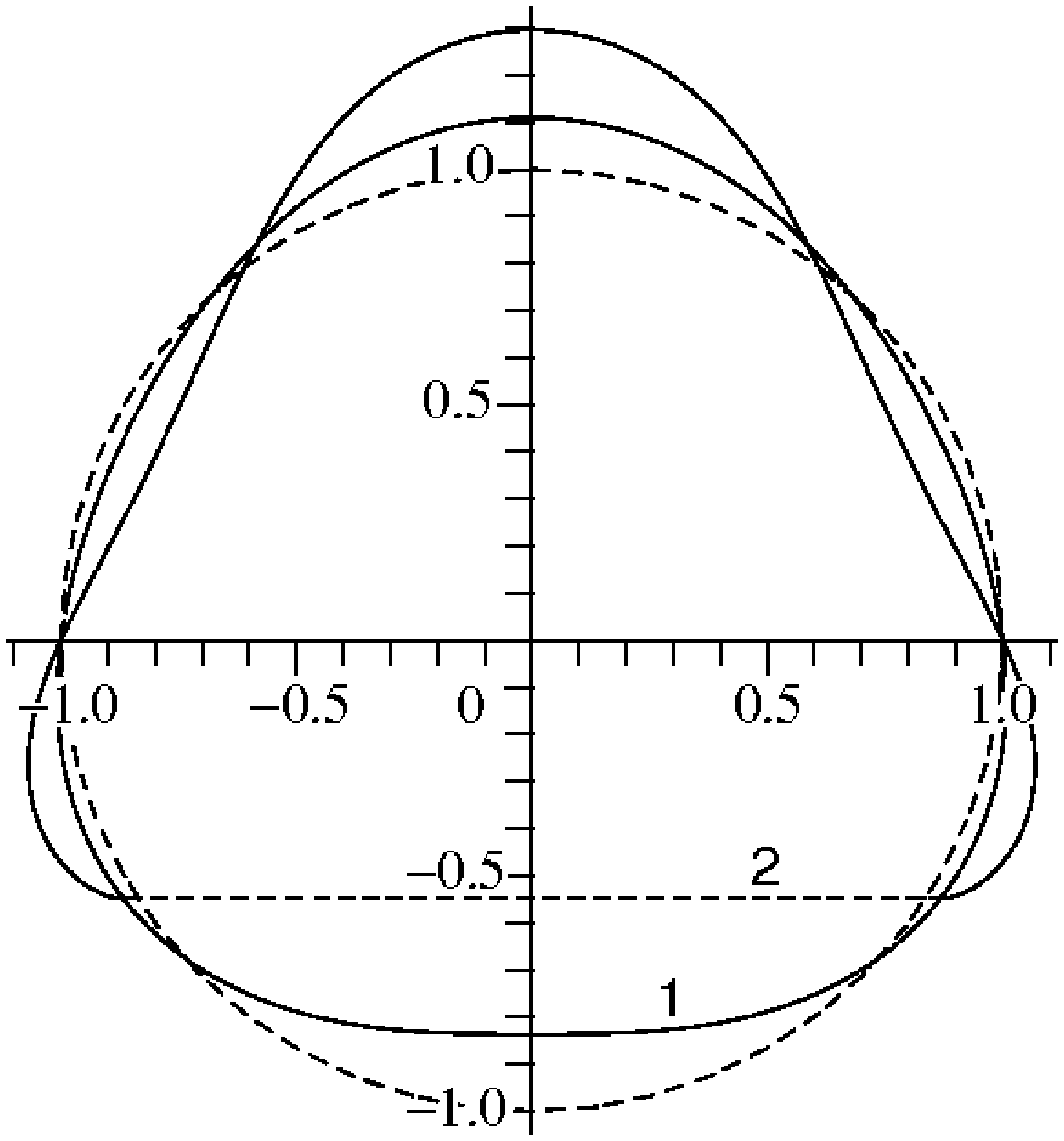}
\nonumber\\
& {\bf a} &\hspace{1.9cm}{\bf b}\nonumber
\ea
\vspace{-0.5cm}
\caption{The shape of the horizon surface of the distorted black hole.
The embedding diagram for the horizon surface can be obtained by
 rotation of the curves on the plots around the vertical axis. The left plot ({\bf a}) shows the
rotation curves for the quadrupole distortion of $a_{2}=1/12$ (line
1), and $a_{2}=-1/12$ (line 2). The right plot ({\bf b}) shows the
rotation curves for the octupole distortion of $a_{1}=-a_{3}=1/20$
(line 1), and $a_{1}=-a_{3}=1/6$ (line 2). The region of negative Gaussian curvature is schematically illustrated on plot {\bf b} by the dashed horizontal line. The rotation curves for positive octupole moments $a_3$ can be obtained by reflection of the lines on plot {\bf b} with respect to the horizontal axis. Dashed lines on both the plots are round circles of radius 1.}
\label{f5} 
\end{center} 
\end{figure}

To construct the embedding we consider a surface
\be
\rho=\rho(\theta)\hh z=z(\theta)
\ee
in 3D Euclidean space with the metric
\begin{equation}
dl^{2}=dz^{2}+d\rho^2+\rho^2d\phi^2.\label{eq20b}
\end{equation}  
The geometry induced on this surface,
\be
dl^2=(z_{,\theta}^2+\rho_{,\theta}^2)d\theta^2+\rho^2d\phi^2\, ,
\ee
coincides with the horizon surface geometry \eq{eq111} if
\ba
\rho&=&e^{-\uu{+}}\sin\theta\hh
z=\int^{\frac{\pi}{2}}_{\theta} d\theta \, Q,\label{em1}\\
Q^2&=&e^{2u_+}-e^{-2u_+}(\cos\theta-u_{+,\theta}\sin\theta)^2\label{em2}.
\ea
Figure \ref{f5} shows the embedding diagrams of the distorted 
event horizon surface for the quadrupole and octupole distortions.

\subsection{Kretchmann invariant}

The function $\uu{+}(\theta)$, which specifies the geometry of the
horizon surface, uniquely determines the geometry of the black hole
interior. In particular, one can obtain expansion of $\cu$ and
$\hv$  at the vicinity of the horizon (see Appendix~B). The first two
terms of this expansion in the powers of $\psi$ are
\ba
\cu &=&\uu{+}-\frac{1}{4} \uuu{+}
\psi^2+\ldots\, ,\n{horu}\\
\hv &=&2\uu{+}-\frac{1}{2}(\uuu{+} -u^{2}_{+,\theta}
+2\cot \theta u_{+,\theta})\psi^{2}+\ldots\, .\nonumber\\
\n{horv}
\ea
Here and later we use the dots $\ldots$ for the omitted  terms of higher order
in $\psi$. We also defined
\be
\uuu{\pm}(\theta)=\sum_{n\geq 0}(\pm 1)^{n}a_{n}n (n+1)P_{n}(\cos \theta)\, .
\ee
In this approximation the metric near the black hole horizon reads
\be
dS_+^{2}=A_+ dT^2+B_+(d\theta^2-d\psi^2)+C_+d\phi^2\, ,
\ee
\ba\n{appr}
A_+&=&\frac{1}{6}\psi^{2}e^{2\uu{+}}
[6-(3\uuu{+}-1)\psi^{2}+\ldots ]\, ,\nonumber\\
B_+&=&\frac{1}{2}e^{2{u_{+}}} [2-(\uuu{+}
+4\cot\theta u_{+,\theta}\\
&-&2u^{2}_{+,\theta}+1)\psi^{2}+\ldots]\, ,\nonumber\\
C_+&=&\frac{1}{2}e^{-2{u_{+}}}\sin^{2}\theta [2+(\uuu{+}
-1)\psi^{2}+\ldots].\nonumber
\ea
This expansion, for example, can be used to determine the value of the
Kretchmann scalar ${\cal K}=R_{\alpha\beta\gamma\delta}
R^{\alpha\beta\gamma\delta}$ at the horizon surface. 
Using \eq{appr} calculations give
\be
{\cal K}_+=12e^{-4u_{+}}
\left(1+u_{+,\theta \theta}
+3\cot\theta u_{+,\theta}-2u^{2}_{+,\theta}\right)^{2}.
\label{eq20}
\ee

\begin{figure}[htb]
\begin{center} 
\HR
\ba
&\includegraphics[height=3.96cm,width=4cm]{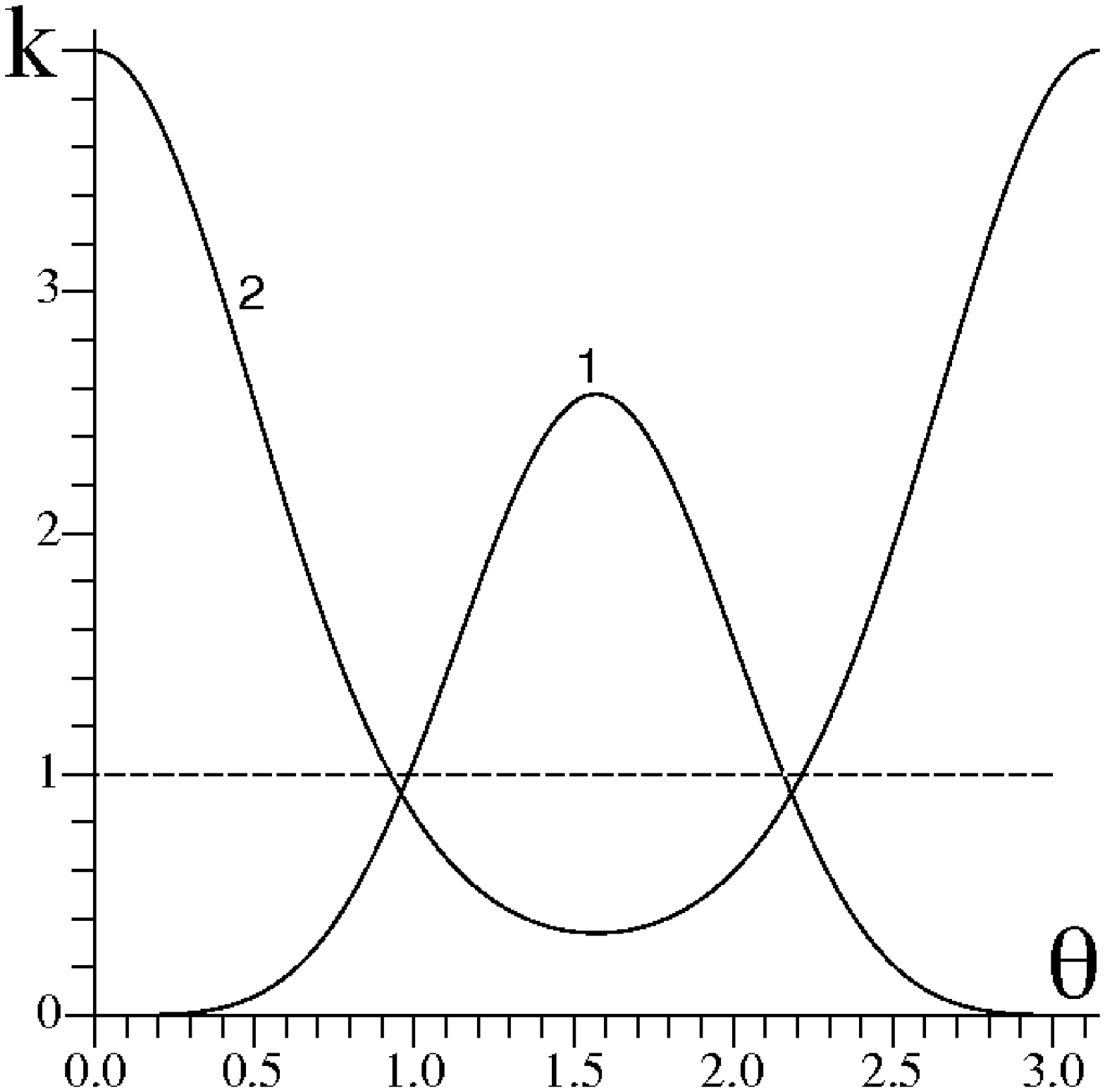}
&
\includegraphics[height=3.96cm,width=4cm]{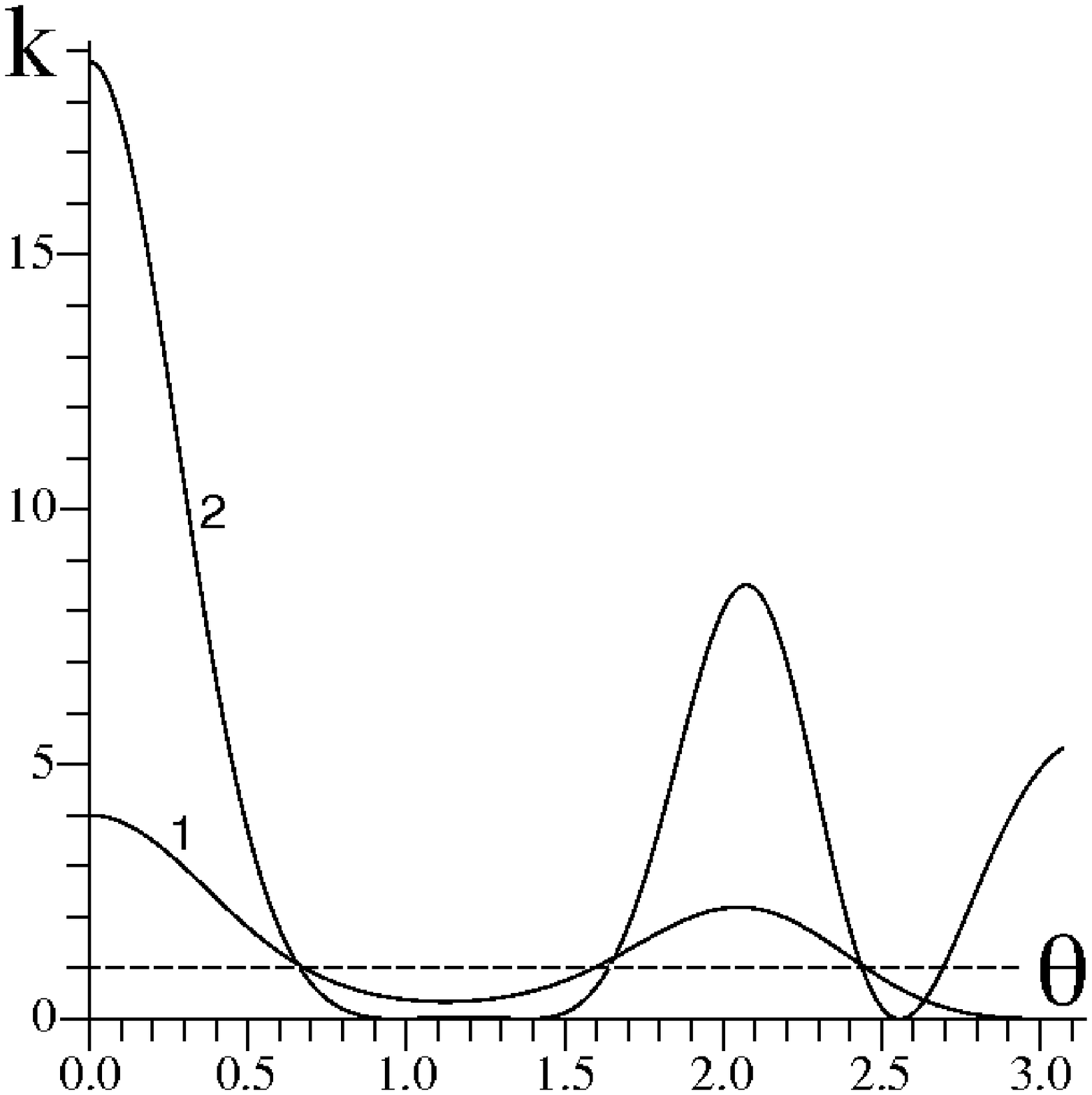}
\nonumber\\
&\hspace{0.1cm} {\bf a} &\hspace{2cm}{\bf b}\nonumber
\ea
\vspace{-0.5cm}
\caption{The ratio $k={\cal K}_+/{\cal K}_{{\ind{Sch}}_{,+}}$, of the
Kretchmann scalar ${\cal K}_+$ on the horizon of the distorted black hole to its undistorted value $k={\cal K}_{{\ind{Sch}}_{,+}}$. Curves on plot ({\bf a}) show $k$ for the quadrupole distortion of
$a_{2}=1/12$ (line 1), and $a_{2}=-1/12$ (line 2). Similar curves on
plot ({\bf b}) show $k$ for the octupole distortion of
$a_{1}=-a_{3}=1/20$ (line 1), and $a_{1}=-a_{3}=1/6$ (line 2). The
dashed horizontal lines at  $k=1$ correspond to the Schwarzschild
black hole.}
\label{f6} 
\end{center} 
\end{figure}

For undistorted Schwarzschild black hole ${\cal
K}_{\ind{Sch}}=12$. Figure \ref{f6} illustrates the ratio, $k={\cal K}_+/{\cal K}_{{\ind{Sch}}_{,+}}$, of the Kretchmann scalars of the distorted and Schwarzschild black holes.
 
Comparing this result with (\ref{eq19c}) one can arrive to the
following relation valid on the horizon surface of a distorted black hole
\be\n{KKK}
{\cal K}_+=12 K_+^2\, .
\ee
This relation is valid not only for axisymmetric case but also for an
arbitrary static distorted black hole (see Appendix \ref{A}).
It is possible to show that the metric of the distorted spherical black holes is of type
D both on the horizon and on the axis of azimuthal symmetry, \cite{Papa}.

\section{Spacetime near the singularity of a distorted black hole}

\subsection{Asymptotic form of the metric}

Because of the symmetry property \eq{symU}, the 
asymptotic form of $\hu$ near the singularity $\psi=\pi$ can
be easily obtained from its asymptotic expansion near the horizon
$\psi=0$. After this, the relation \eq{intv} allows one to find the
expansion of $\hv$. The expansions for $\hu$ and $\hv$ near the
singularity are given in Appendix~B. Using these expansions one
can obtain the asymptotic form of the metric \eq{mm} at
$\psi_-=\pi-\psi \to 0$
\be\n{msing}
dS_-^2=A_-dT^2+B_-(d\theta^2-d\psi_-^2)+C_- d\phi^2\, ,
\ee
\ba\n{msing1}
A_-&=&\frac{8e^{2u_-}}{3\psi_-^2} [6
+\left(3u_{-,\theta\theta}+
3\cot\theta u_{-,\theta}-1\right)\psi_-^2\ldots ]\, ,\nonumber\\
B_-&=&\frac{e^{-6u_-}}{96} \psi_-^4 [6-(9u_{-,\theta\theta}\\
&-& 3\cot\theta u_{-,\theta}-6(u_{-,\theta})^2+1)\psi_-^2 +\ldots ]\, ,\nonumber\\
C_-&=&\frac{e^{-2u_-}}{96} \sin^2\theta \psi_-^4\nonumber\\
&\times& [6-(3u_{-,\theta\theta}
+3\cot\theta u_{-,\theta}+1 )\psi_-^2\ldots ]\, .\nonumber
\ea
Expressions \eq{msing1} are sufficient to calculate the Kretchmann
scalar near the singularity up to the second order in $\psi_-$
corrections
\ba
{\cal K}_-&=&\frac{49152\,e^{12u_-}}{\psi_-^{12}}\left (1+\frac{1}{2}\psi_-^2 \tilde{{\cal K}}_-^{(2)}+\ldots \right)\, ,\n{kre}\\
\tilde{{\cal K}}_-^{(2)}&=&1+3u_{-,\theta\theta}
-6 (u_{-,\theta})^2-3\cot\theta u_{-,\theta}\, .\n{kre2}
\ea
Higher order terms can be obtained by using the relations given in Appendix~B. 
In the absence of distortion, when $u_-=0$, the Kretchmann scalar
does not depend on $\theta$
\be
{\cal K}_{{\ind{Sch}}_{,-}}=\frac{49152}{\psi_-^{12}}\, .
\ee
This is the value of ${\cal K}_-$ for the Schwarzschild geometry. It
can be shown that  the metric of a distorted black hole is of type D
near the  singularity \cite{Papa}.

\subsection{Stretched singularity}

For the Schwarzschild geometry the metric near the singularity
\be
dS_{-}^2\approx -\frac{1}{16}\psi_-^4 d\psi_-^2+ 
\frac{16}{\psi_-^2} dT^2+\frac{\psi_-^4}{16}d\omega^2\, ,
\ee
can be written in the form
\be\n{kasn}
dS_{-}^2\approx -d\tau^2+ \frac{16}{(12\tau)^{2/3}}
dT^2+\frac{(12\tau)^{4/3}}{16}d\omega^2\, .
\ee
Here $\tau=-\psi_-^3/12$ is the proper time of a free fall to the
singularity along the geodesic  $T,\theta,\phi=const$. The quantity
$\tau$ is negative, and it reaches $0$ at the singularity.  The
metric \eq{kasn} has the Kasner-like behavior with indices
$(-1/3,2/3,2/3)$. It describes a metric of collapsing anisotropic
universe, that shrinks in $\theta$-$\phi$ directions, and
expands in $T-$ direction.

The Kretchmann invariant as a function of the proper time has the
following asymptotic form
\be\n{KSch}
{\cal K}_{{\ind{Sch}}_{,-}}\approx \frac{64}{27\tau^4}\, .
\ee
This relation shows that the surface of constant ${\cal K}_{{\ind{Sch}}_{,-}}$ is at the same time a surface of constant $\tau$.

Spacetime in the region where the curvature is of order of the
Planckian curvature requires quantum gravity for its description.
For the Schwarzschild geometry at the surface  where ${\cal K}_{{\ind{Sch}}_{,-}}\sim l_{Pl}^{-4}$ the proper time $\tau$ is of order of the
Planckian time $\tau_{Pl}$.  Since one cannot rely on the classical
description in this domain, it is natural to cut the region where the
curvature is higher than the Planckian one and to consider its
boundary as the {\em stretched} or {\em `physical singularity'}.  For the
Schwarzschild metric the stretched singularity surface has the
topology $R^1\times S^2$. Its metric is a direct sum of the metric of
a round two-sphere and a line.

What happens to the stretched singularity when the metric of the
black hole is distorted? To answer this question we use the
asymptotic form of the metric near the singularity, \eq{msing}.
Consider a timelike geodesic lying on the `plane' $T=const$,
$\phi=const$. We call such a geodesic `radial'. It can be shown (see Appendix C) that a 'radial' geodesic is uniquely determined by the limiting value $\vartheta$ of its angular parameter $\theta$ at which it crosses the singularity. Denote by $\tau$ the proper time along the 'radial' geodesic to its end point at the singularity.
In coordinates  $(\tau,\vartheta)$ the metric $dS_{-}^2$ is given by \eq{kasn} where $d\omega^2$ is replaced by
\be
d\sigma^2_-=e^{-2u_-} d\theta^2+e^{2u_-}\sin^2\theta d\phi^2\, .\n{dsm}
\ee
We can use $(\tau,\vartheta)$ as new coordinates in the vicinity of the
singularity. Relations \eq{pt}, \eq{tt} connect these `new'
coordinates with the `old' ones $(\psi_-,\theta)$. The Kretchmann
scalar \eq{kre}, \eq{kre2} in `new' coordinates reads
\ba\n{kret}
{\cal K}_{-}&=&\frac{64}{27\tau^4} [1+{\cal K}_{-}^{(2)}\tau^{2/3}+O(\tau^{4/3})]\, ,\\
{\cal K}_{-}^{(2)}&=&\frac{1}{2}(12)^{2/3}e^{2u_-(\vartheta)} [1
+3u_{-,\vartheta\vartheta}\nonumber\\
&-&6 (u_{-,\vartheta})^2-3\cot\theta u_{-,\vartheta}]\label{kret2} .
\ea
The expansion \eq{kret} coincides in the leading order with \eq{KSch}.
Hence, in the presence of distortion surfaces of equal
${\cal K}$ are again (in the leading order) surfaces of
constant $\tau$.

\subsection{Shape of equi-curvature surfaces}

A surface $\Sigma_{-}$ where the  Kretchmann scalar has
constant value ${\cal K}_{-}={\cal K}_0$.  In the vicinity of the
singularity  (in the leading order in $\psi_-$) $\psi_-$ and $\theta$
on $\Sigma_-$ are related as follows
\be
\psi_-=\kappa_- e^{u_-}\hh
\kappa_-=(49152/{\cal K}_0)^{1/12}\,,
\ee
and one has the relation $\psi_{-,\theta}=\psi_- u_{-,\theta}$. Consider the
induced geometry on $\Sigma_{-}$. Using this relation one can conclude
that $d\psi_-^2$ term in \eq{msing} gives quadratic in $\psi_-$
corrections only. Neglecting all such terms in \eq{msing} we obtain the
following expression for the leading asymptotic for the induced
metric $dl_{-}^2$ on $\Sigma_{-}$
\be
dl_{-}^2\approx\frac{16}{\kappa_-^2}dT^2+\frac{\kappa_-^4}{16} d\sigma^2_-\, ,
\ee
where $d\sigma^2_-$ is given by \eq{dsm}. The surface $\Sigma_{-}$ has the same topology $R^1\times S^2$ as in the absence of distortion, but its geometry is different. This
difference manifests itself in the shape of $T=const$ 2D
surfaces. The information about the shape is encoded in the 2D metric
$d\sigma^2_-$. The total area of the surface is $4\pi$. The metric \eq{dsm} can be obtained from the
horizon metric $d\sigma^2_+$, \eq{dsp}, by a simple change $u_+\to
-u_-$. Under this transformation the even multipole coefficients are
invariant, while the odd coefficients change their sign (see \eq{u}).
In particular, the embedding diagrams for the metric \eq{dsm} are
the same as shown on Figure \ref{f5} ({\bf a}), with the same value of $a_2$, and with the opposite value of the octupole moment $a_3$, as shown on Figure \ref{f5} ({\bf b}).

\section{Exact solutions}

For any given set of multipole coefficients $a_i$ that
determine $\hu$, the corresponding function $\hv$ can be found
explicitly in terms of elementary functions. Since the general
expression for $\hv$ is rather cumbersome we do not  present it here.
Instead, we consider special case of the quadrupole and octupole distortions when the gravitational potential 
$\cu$ is of the form
\be
{\cal U}=-\frac{3}{2}a_2{\cal P}_{3/2}-
\frac{5}{2}a_3\cos\psi\cos\theta{\cal P}_{5/2},
\ee
where
\be
{\cal P}_q=\sin^2\psi+\sin^2\theta-q\sin^2\psi\sin^2\theta.
\ee
Substituting this solution into
equations \eq{eq3a}, \eq{eq3b}, and integrating according to equation
\eq{intv} we derive
\ba
\hat{V}&=&\sin^2\theta [-3a_2\cos\psi+
\frac{5}{2}a_3\cos\theta (1-3\cos^2\psi)]\nonumber\\ 
&+&\frac{1}{2}\sin^2\psi \sin^2\theta [a_2^2 {\cal V}_{22}
+2a_2 a_3 {\cal V}_{23}+a_3^2 {\cal V}_{33}],
\ea
where
\ba
{\cal V}_{22}&=&9[1-{\cal P}_{9/8}]\, ,\nonumber\\ 
{\cal V}_{23}&=&15\cos\psi\cos\theta [1-3/2{\cal P}_{3/2}]\, ,\\ 
{\cal V}_{33}&=&\frac{25}{4} [4-12{\cal P}_{39/24}-18{\cal P}^2_2+27{\cal P}^2_{11/6}]\, .\nonumber
\ea
Exterior metric for a black hole distorted by a quadrupole field was derived in \cite{Dor}.
 
\begin{figure}[htb]
\begin{center}
\HR
\ba
&\includegraphics[height=3.88cm,width=4cm]{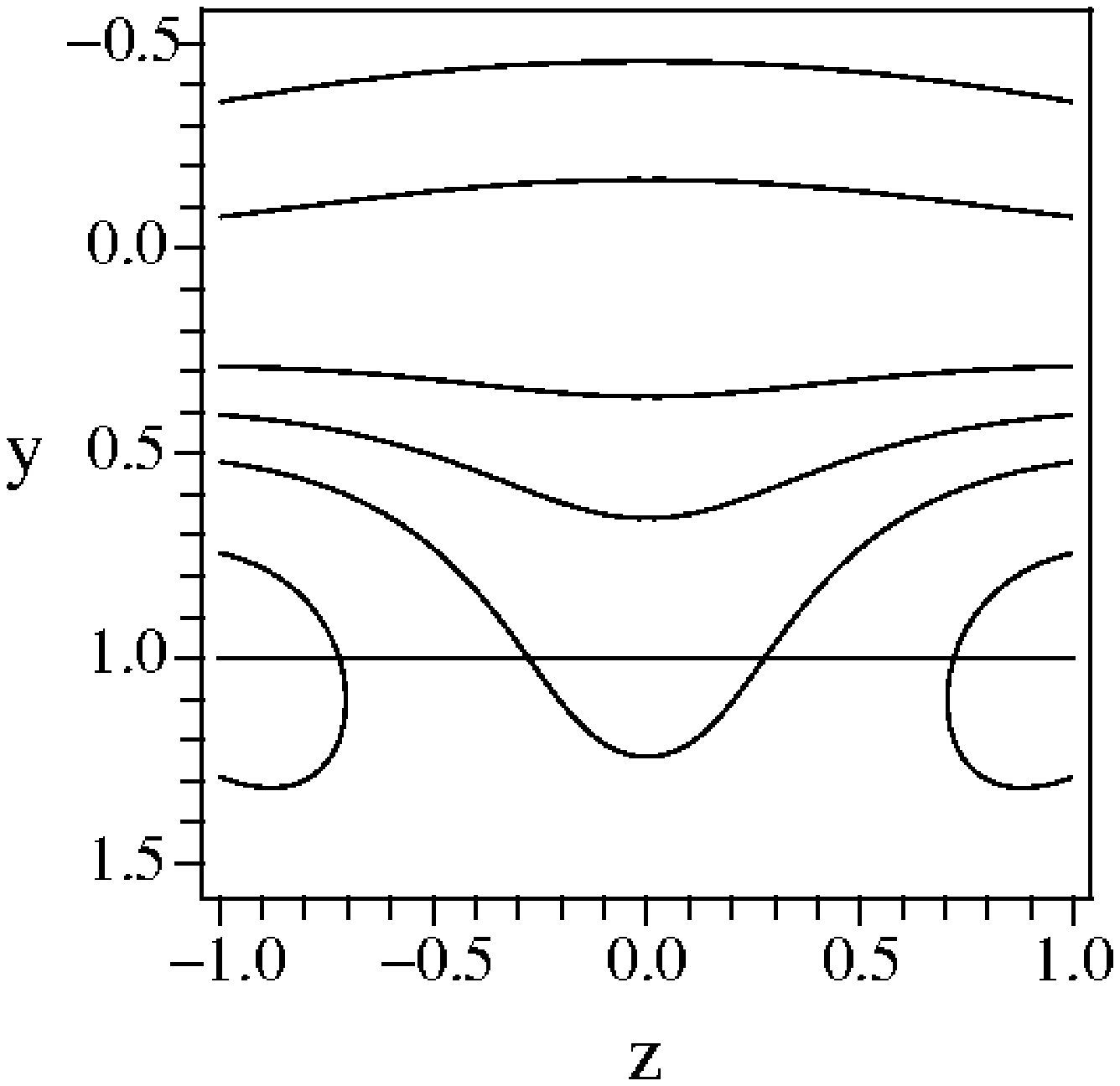}
&\includegraphics[height=3.91cm,width=4cm]{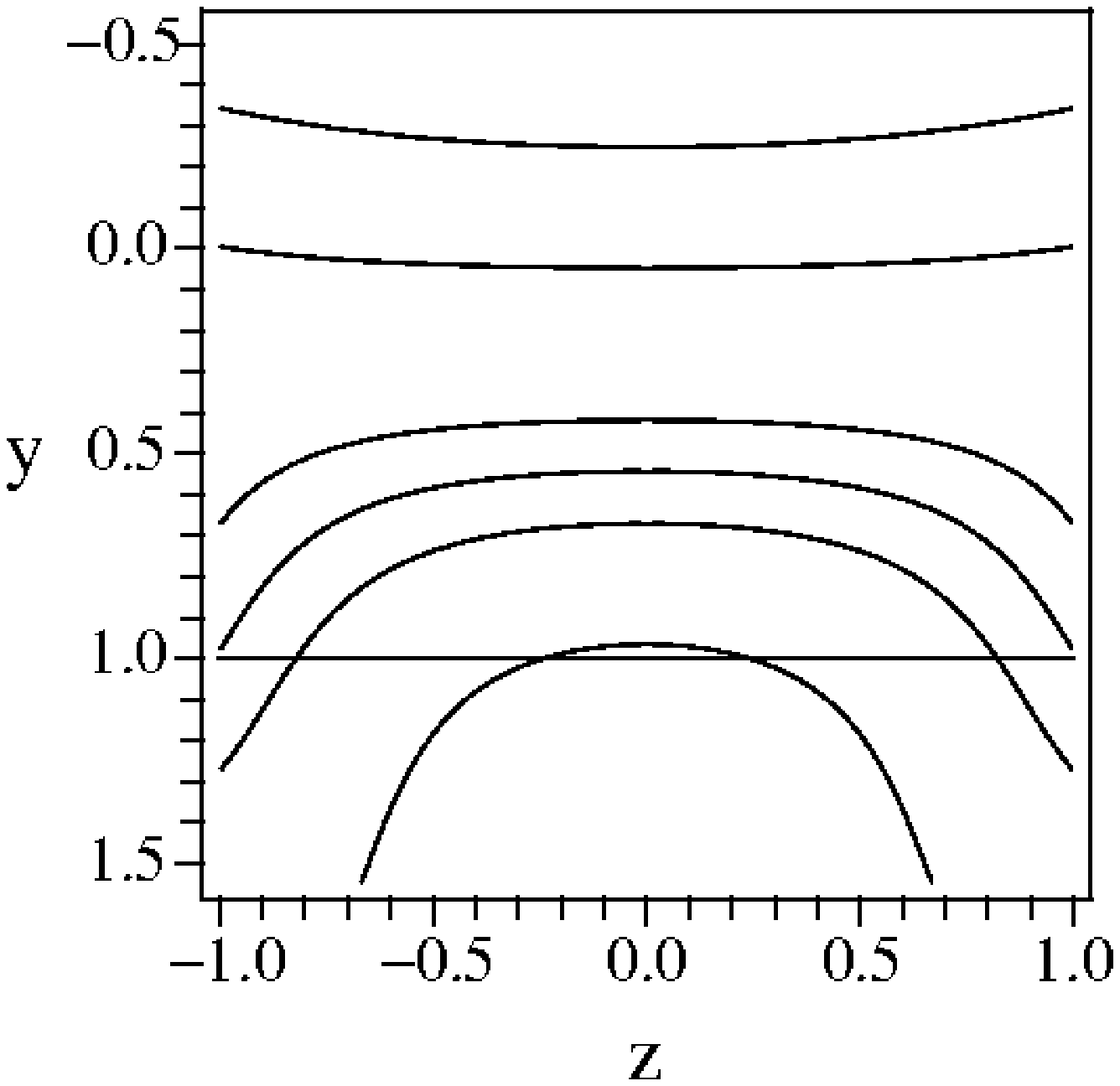}
\nonumber\vspace{-2cm}\\
&\hspace{0.5cm} {\bf a} &\hspace{2.15cm}{\bf b}\nonumber
\ea
\vspace{-0.8cm}
\caption{The contour lines of ${\cal K}$ for the quadrupole distortion of $a_{2}=1/12$ ({\bf a}), and $a_{2}=-1/12$ ({\bf b}). The horizontal line $y=1$ represents the event horizon.} \label{f7} 
\end{center} 
\end{figure}

\begin{figure}[htb]
\begin{center} 
\HR
\ba
&\includegraphics[height=3.92cm,width=4cm]{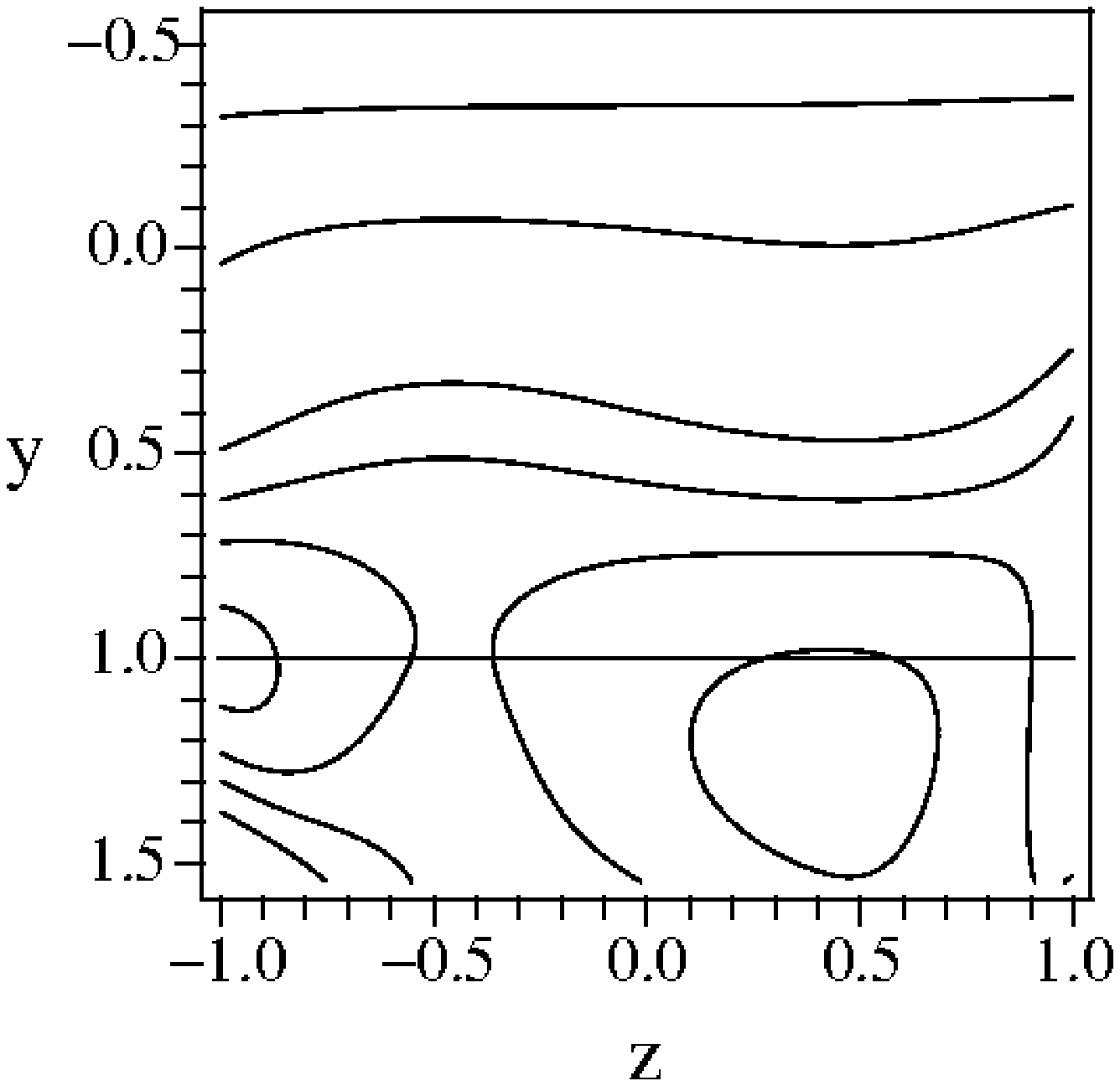}
&\includegraphics[height=3.91cm,width=4cm]{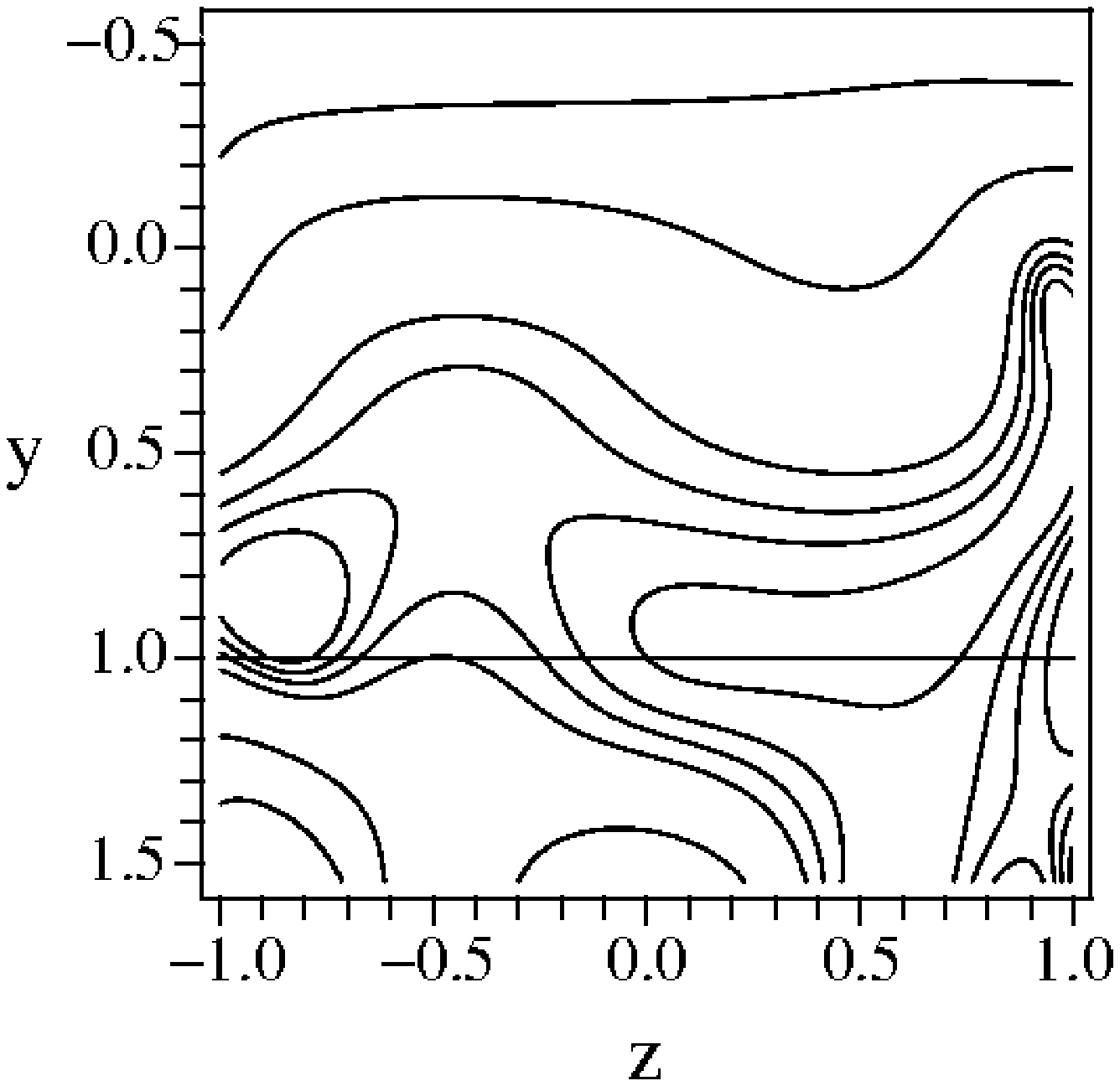}
\nonumber\\ 
& \hspace{0.5cm}{\bf a} &\hspace{2.15cm}{\bf b}\nonumber
\ea
\vspace{-0.8cm}
\caption{The contour lines of ${\cal K}$ for the octupole distortion of $a_{1}=-a_{3}=1/20$ ({\bf a}), and $a_{1}=-a_{3}=1/6$ ({\bf b}). The horizontal line $y=1$ represents the event horizon.} 
\label{f8} 
\end{center} 
\end{figure}

Using GRtensorII package we calculated the Kretchmann  scalar  ${\cal K}$
for this distortion. We made the calculations for both the inner
and external regions. It is easy to check that ${\cal K}$'s in these
regions are related by the analytical continuation \eq{eq1a}. Figures \ref{f7} and \ref{f8} show the
contour lines of ${\cal K}$ for the quadrupole and octupole
distortions respectively. In order to plot both the exterior and
interior regions simultaneously we introduce new coordinate 
\ba
y= \begin{cases}
\cos\psi \, ,& \mbox{for   } \psi\in (0,\pi)\, ;\nonumber\\
{}&\\ 
\cosh \tilde{\psi}\, , &\mbox{for   } \psi\in (0,\infty)\, .\nonumber\\
\end{cases}
\ea
The sector $y\in (-1,1)$, $z\in[-1,1]$, where $z=\cos\theta$, covers the inner region, and the sector
$y\in (1,\infty)$, $z\in[-1,1]$ covers the exterior of the black hole. 

\section{Interior of a caged black hole}
\n{cbh}

In this Section we apply the obtained results to a special case of the
so called `caged' black hole. Such a black hole is a solution of the
vacuum Einstein equations for a spacetime where one (or more) spatial
dimensions are compactified. Caged black holes in 4D spacetime were
discussed in \cite{FrFr} (see also \cite{Myers:87}, \cite{BoPe:90}). As the
result of the compactification the event horizon of the black hole is
distorted. The metric is axisymmetric and is a special case of the
Weyl solution. The value of $\hu$ at the horizon found in \cite{FrFr}
[ equation (62)] gives
\be
\hat{U}(0,\theta)=\frac{\mu}{\pi}\, \ln(4\pi)+\frac{1}{2}\, \ln \left[
f\left(\frac{\mu+z}{2}\right)f\left(\frac{\mu-z}{2}\right)\right],\n{29}
\ee
where $|z|\le \mu$ and $0<\mu<\pi$. In this case $z=\mu \cos\theta$, $\mu$ is a dimensionless parameter equal to
the ratio of the black hole mass, $m$, to the radius of the
compactification, $L$, $\mu=m/L$, and
\be
f(x)=\frac{1}{\pi^2}\, x\, \sin x\, \Gamma^2(x/\pi).\label{30}
\ee
The function $f(x)$ has the following properties:
\be
f(0)=1\, ,\hspace{1cm}f(\pi/2)=\frac{1}{2} \, ,\hspace{1cm}f(\pi)=0,\n{31}
\ee
and at the interval $0\le x \le \pi$ it can be approximated by a
linear function
\be
f(x)\approx 1-\frac{x}{\pi}\n{32} 
\ee
with accuracy of 1\% .
The coefficients $a_n$ for this solution $\hu$ can be obtained from \eq{an}. Let us emphasize that since the function $\hat{U}(0,\theta)$ is
invariant under the transformation $\theta\rightarrow\pi-\theta$, the
coefficients $a_n$ for odd $n$ vanish (see equations \eq{u}, \eq{f2a}). This implies that
$\uu{-}=\uu{+}$ and the boundary value of $\hu$ at $\psi=\pi$
coincides with the boundary value of this function at the horizon,
$\psi=0$ (see equation \eq{eqf2b}), and we have (see equation \eq{eq4b})
\begin{equation}
u_{0}=\frac{\mu}{\pi}\, \ln(4\pi)+\frac{1}{2}\, \ln [f(\mu)].\n{uo}
\end{equation}
\begin{figure}[htb]
\begin{center}
\HR
\ba
&\includegraphics[height=4cm,width=2.649cm]{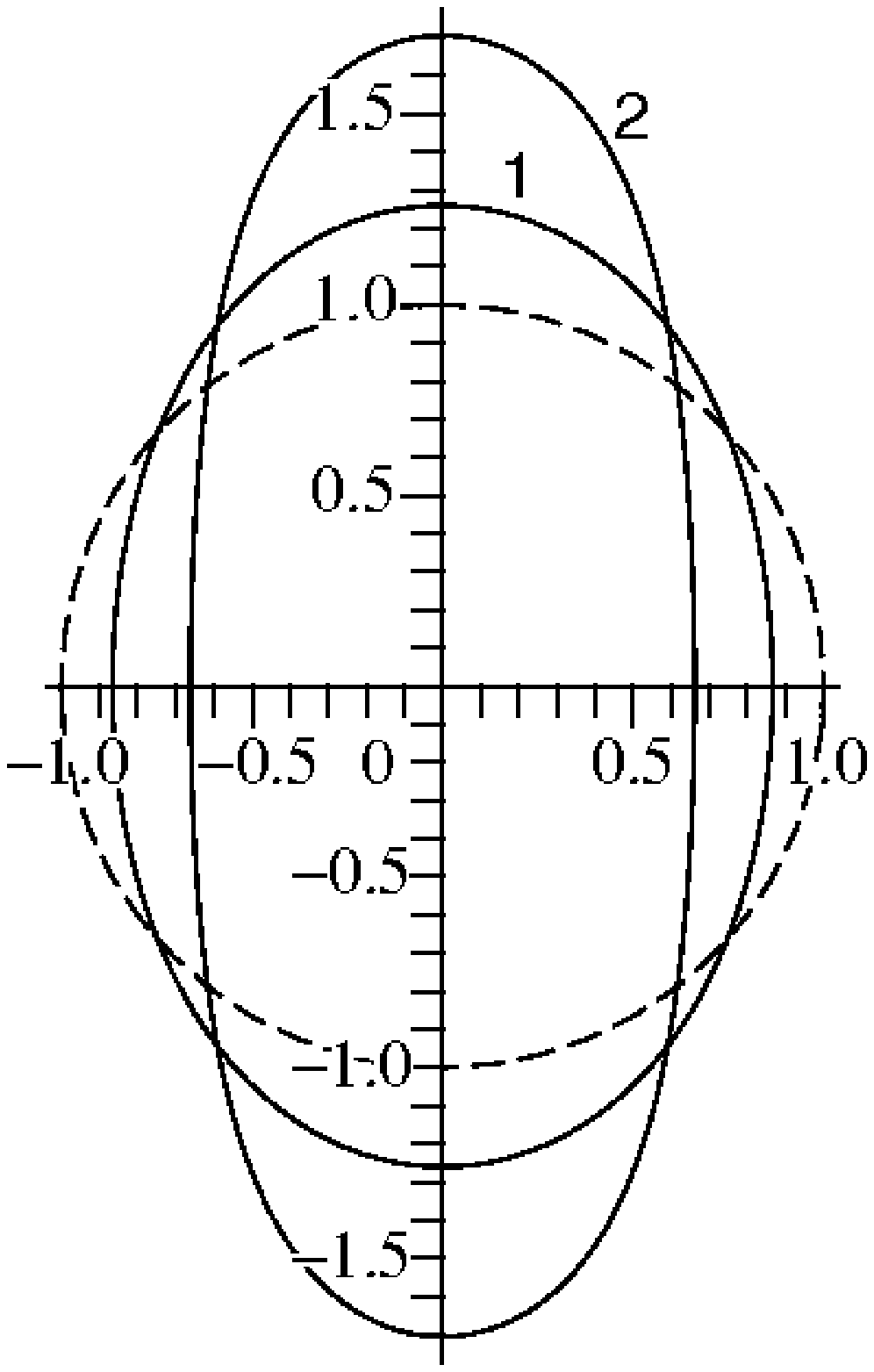}
\hspace{1cm}
&\includegraphics[height=3.96cm,width=4cm]{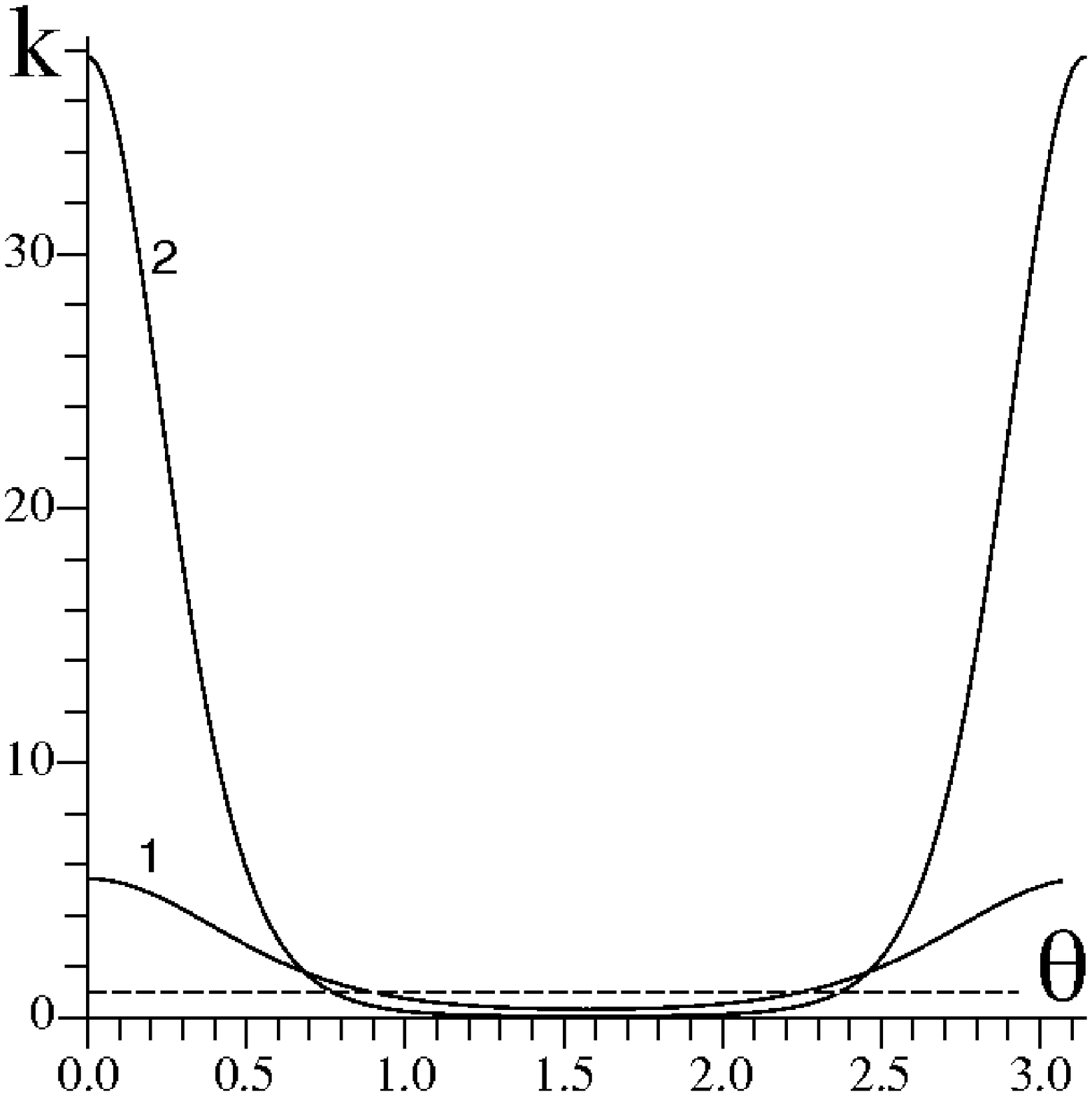}
\nonumber\\ 
& \hspace{-1cm}{\bf a} &\hspace{2cm}{\bf b}\nonumber
\ea
\vspace{-0.8cm}
\caption{
({\bf a}) The shape of the distorted event horizon surface for the caged
black hole. (1): $\mu=2/3\pi$, (2): $\mu=6/7\pi$. The dashed
circle corresponds to the Schwarzschild black hole. The embedding
surface is obtained by rotation of these curves around the
vertical axis. ({\bf b})  The Kretchmann scalars ratio $k={\cal
K}_+/{\cal K}_{{\ind{Sch}}_{,+}}$ on the horizon for the same values of $\mu$
(lines 1 and 2). The dashed horizontal line corresponds to the
Schwarzschild black hole.} \label{f9} 
\end{center} 
\end{figure}

\begin{figure}[htb]
\begin{center}
\includegraphics[height=3.38cm,width=5cm]{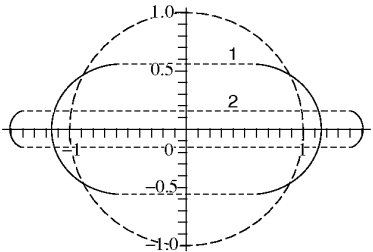}
\caption{The shape of the distorted {\em `physical singularity'} for the
caged black hole of $\mu=2/3\pi$ (line 1), and $\mu=6/7\pi$
(line 2).  The region of negative Gaussian curvature is schematically illustrated by the dashed horizontal lines. The dashed circle corresponds to the Schwarzschild black
hole. The embedding surface is obtained by rotation of these
curves around the vertical axis.} \label{f10}  
\end{center} 
\end{figure}
From equations \eq{29} and \eq{uo} we derive
\be
u(\theta)=\uu{\pm}(\theta)\approx\frac{1}{2}\ln\left(\frac{4\pi(\pi-\mu)+\mu^{2}\sin^{2}\theta}{4\pi(\pi-\mu)}\right).
\n{33}
\ee
The metric \eq{dsp} on the surface of the horizon is
\ba
d\sigma^2_+&\approx& \left(1+\frac{\mu^2\sin^2\theta}{4\pi(\pi-\mu)}\right)d\theta^2\nonumber\\
&+&\left(1+\frac{\mu^2\sin^2\theta}{4\pi(\pi-\mu)}\right)^{-1}\sin^2\theta\, d\phi^2\, .
\ea
Using equations \eq{eq20}, \eq{KKK}, and \eq{kret}, \eq{kret2} we can calculate the Kretchmann scalar at the horizon surface of a caged black hole
\be
{\cal K}_+\approx12\left(4\pi(\pi-\mu)\right)^{4}\frac{(4\pi(\pi-\mu)+\mu^{2}+3\mu^{2}\cos^{2}\theta)^{2}}{\left(4\pi(\pi-\mu)+\mu^{2}\sin^{2}\theta\right)^{6}},\n{34}
\ee
and in the vicinity of its {\em 'physical singularity'}
\ba\n{kretcage}
{\cal K}_-&\approx&\frac{64}{27\tau^4} [1+{\cal K}^{(2)}\tau^{2/3}+O(\tau^{4/3})]\, ,\\
{\cal K}_-^{(2)}&\approx&\frac{1}{2}(12)^{2/3}\bigl[\left(8\pi(\pi-\mu)-\mu^{2}\sin^{2}\theta\right)^{2}\bigr.\nonumber\\
&+&\bigl.3\mu^4\sin^2\theta(13\sin^2\theta-16)\bigr]\nonumber\\
&\times&\left[16\pi(\pi-\mu)\left(4\pi(\pi-\mu)+\mu^{2}\sin^{2}\theta\right)\right]^{-1}\label{kret2cage} .
\ea

Applying equations \eq{em1}, \eq{em2}, and the transformation $u_+\to
-u_-$ respectively we can construct embedding of these surfaces. Figure \ref{f9}
illustrates the shapes of the distorted event horizon surface and the ratio of the Kretchmann scalars , $k={\cal K}_+/{\cal K}_{{\ind{Sch}}_{,+}}$, of the caged and Schwarzschild black holes. 
The metric \eq{dsm} on the surface of the {\em stretched singularity} is
\ba
d\sigma^2_-&\approx& \left(1+\frac{\mu^2\sin^2\theta}{4\pi(\pi-\mu)}\right)^{-1}d\theta^2\nonumber\\
&+&\left(1+\frac{\mu^2\sin^2\theta}{4\pi(\pi-\mu)}\right)\sin^2\theta\, d\phi^2\, .
\ea
The shape of distorted {\em 'physical singularity'} is illustrated on Figure \ref{f10}.

\section{Discussions}

Let us summarize the obtained results. We considered the geometry of
static vacuum axisymmetric distorted black holes. We focused mainly
on the properties of the horizon and interior of such black holes. The
geometry of a distorted black hole is uniquely determined by the
'gravitational potential' $U$ which is a solution of the 3D flat
Laplace (in the exterior) or d'Alembert (in the interior) equation.
After solving this `master' equation, the second function $V$, which
enters the metric, can be obtained by a simple integration.

The 'gravitational potential' $U$ can be written as a superposition of
the Schwarzschild potential $U_{\ind{Sch}}$ and the distortion potential $\hu$. The distortion $\hu$ is determined by the values of the multipole
coefficients $a_i$ obeying the constraints \eq{eq7b}. The distortion
potential in the black hole interior possesses a remarkable
discrete symmetry \eq{symU} which relates the value of  $\hu$ in the
vicinity of the singularity to its value at the horizon. Thus, the
functions $u_{\pm}(\theta)$, (see \eq{u}), determine both, the shape of the
horizon and the leading asymptotics of the metric and curvature
invariants near the singularity. 

Qualitatively, the shape of the event horizon surface of distorted black
hole is similar to the shape of equipotential surfaces in the linearized
(Newtonian) gravity. Namely, consider a point-like mass $M$. In the presence of a
quadrupole distortion its Newtonian gravitational potential reads (up to constant $a_0$)
\be\n{pot}
\Phi=-\frac{M}{R}+\Delta \Phi \hh \Delta\Phi=\frac{a_2}{2}R^2(3\cos^2\theta-1)\, ,
\ee  
where $R=\sqrt{x^2+y^2+z^2}$ is the radial distance from the mass $M$, and
$a_2$ is the value of the quadrupole moment. For positive $a_2$ such a
distortion is generated, for example, by a ring of mass $m$ and radius $d\gg R$ located in
the equatorial plane. For such a ring $a_2=m/(2d^3)$.
Similarly, a negative $a_2$ is generated, for example, by two point
masses $m$ located on the axis of symmetry on the opposite sides of
the mass $M$ at the distance $d\gg R$. In this case $a_2=-2m/d^3$. We
consider $R\sim M$ and assume that the distortion $\Delta\Phi$ is small. Then the change $\delta R$ in the position of the equipotential surface for \eq{pot} with respect to the position of the
unperturbed surface of $R_0=const$ is
\be
\delta R=-\frac{a_2 R_0^4}{2M}(3\cos^2\theta-1)\, .
\ee
Thus the quadrupole distortion deforms the equipotential surfaces and
makes them either oblate (for $a_2>0$), or prolate (for $a_2<0$). This
property is similar to the property of the horizon surface for the
distorted black hole (see Figure \ref{f5} ({\bf a})).

It should be emphasized that the linear approximation is not
sufficient for the `explanation' of the Kretchmann invariant
properties. Really, in the linear approximation
\be
ds^2=-(1+2\Phi)dt^2+(1-2\Phi)(dx^2+dy^2+dz^2)\, ,
\ee
the Kretchmann scalar is
\be\n{KrN}
{\cal K}\approx 8\Phi_{,ij}\Phi^{,ij}=48\left[a_2^2-\frac{M(M+2R\Phi)}{R^6}\right].
\ee
Its variation under the small distortion $\Delta\Phi$ is
\be
\delta {\cal K}=-\frac{2M^2}{R_0^7}\delta R\,.
\ee
Hence in the weak field approximation, ${\cal K}$ is larger at the
points where $\delta R<0$, such as at the pole of the oblate
equipotential surface (for $a_2>0$), and in the equatorial points of
the prolate surface (for $a_2<0$). This behavior of ${\cal K}$ in the
weak field limit is opposite to the behavior of ${\cal K}$ on the
horizon of the distorted black hole (see e.g. \eq{KKK}). This
difference demonstrate that non-linear effects and the spatial
curvature are important near the horizon.

The property \eq{KKK} has an important consequence for caged black
holes discussed in Section~\ref{cbh}. For $\mu$ close to $\pi$, when
the `north' and `south' poles of the caged black hole are close to
one another, the Gaussian curvature (and hence the Kretchmann
invariant) becomes large at the poles. In other words, in the
infinitely slow merger transition, the region of a very high curvature
'leaks' through the horizon in the vicinity of the black hole poles. When
this curvature reaches the Planckian value, one can say that the
{\em 'physical singularity'} (as defined in subsection B of Section~V) becomes naked.
This may indicate that during the phase transition between black-hole and black-string  phases one can expect a formation of a naked {\em 'physical singularity'}. Whether this conclusion remains valid for higher dimensional caged black holes and beyond the adiabatic approximation
is an interesting open question.

\begin{acknowledgments}
This research was supported  by the Natural Sciences and Engineering
Research Council of Canada and by the Killam Trust. 
\end{acknowledgments}

\appendix

\section{\label{A} Kretchmann invariant and Gaussian curvature on the
horizon of a static black hole}

In this appendix we show that the Kretchmann invariant
${\cal K}=R_{\mu\nu\lambda\rho}R^{\mu\nu\lambda\rho}$ calculated at the
horizon of a 4D static  distorted black hole is related to the Gaussian
curvature of the 2D horizon surface $K_0$ as follows
\be\n{res}
{\cal K}=12 K^2_0\, .
\ee
Let us emphasize that this relation is valid for an arbitrary
(not necessarily axisymmetric) distorted black hole. To establish
this relation we use the results of paper \cite{FrSa}.

The metric near the horizon of a static vacuum distorted black hole can be
written in the Israel coordinates  \cite{Israel} as follows 
\be\n{dbh}
ds^2=-X dt^2+\frac{dX^2}{4 \kappa^2 X}+h_{ab} d\theta^a d\theta^b\, ,
\ee
where $-X=\xi_t^2$ is the square of the timelike Killing vector
$\xi_t=\pa_t$,
\be
\kappa=\frac{1}{2}{\cal D}^{1/2}\, ,
{\cal D}= -2\xi_{a|b}\xi^{a|b}=\Box X=X^{-1} (\nabla X)^2\, .
\ee
Here $(\ldots)_{a|b}$ means the covariant derivative with respect to
 2D metric $h_{ab}$, $\nabla$ and $\Box$ are the operators associated with this
metric. Using the properties of the Killing vector one can show that
in the Ricci flat spacetime the following relations are valid
\ba
\pa_X h_{ab}&=&\kappa^{-1}k_{ab}\hhh
\pa_X \kappa=-\frac{1}{2}k\, ,\\
\pa_a\kappa&=&-X(k_{|a}-{k_a^{\,\,b}}_{|b}\hhh  k=h^{ab}k_{ab}\, ,\\
X\pa_X k_a^{\,\,b}&=&-k_a^{\,\,b}+\frac{1}{2}k\delta_a^{\,\,b}-\frac{1}{2}
(\kappa^{-1})_{|a}^{\,\,\,\, |a}\\
&-&(4\kappa)^{-1}X[(k_{cd}k^{cd}-k^2)\delta_a^{\,\,b}+2k k_a^{\,\,b}]\, ,\\
K&=&\kappa k-\frac{1}{2}X (k_{ab}k^{ab}-k^2)\, .
\ea
Here $K$ is the Gaussian curvature of 2D surface $X=const$.  The
Kretchmann invariant for the metric \eq{dbh} can be written as
follows (see equation (4.18) of \cite{FrSa})
\be\n{K}
{\cal K}=8\kappa^2[k_{ab}k^{ab}
+k^2+2(X\kappa^2)^{-1}\kappa_{|a}\kappa^{|a}]\, .
\ee
One can show that quantities $h_{ab}$, $k_{ab}$ and $\kappa$ are
finite at the horizon and its vicinity they allow a representation in
the form of the Taylor series 
\ba
\kappa&=&\kappa_0+\kappa_1 X+O(X^2)\hhh \kappa_{0|a}=0\, ,\n{k0}\\
k_{a}^{\,\,b}&=&\frac{1}{2}k_0\delta_a^{\,\,b}+O(X)\, ,\n{kab}\\
K&=&K_0+O(X)\hhh K_0=\kappa_0 k_0\, .\n{K0}
\ea
Equations \eq{k0} imply that the third term in the squared brackets of
\eq{K} vanish at the horizon $X=0$. Simplifying the other two terms by
using \eq{kab} and \eq{K0} one obtains \eq{res}.

\section{\label{B} Asymptotic expansions near the horizon and
singularity}

The solution \eq{eq6} can be used to find the asymptotic behavior of
${\cal U}$ near the horizon, $\psi=0$,  and singularity, $\psi=\pi$. 
To deal with both cases simultaneously, we denote $\psi_{+}=\psi$ and
$\psi_-=\pi-\psi$. The function ${\cal U}$ is an even function of
$\psi_{\sigma}$ \ ($\sigma=\pm$), and it has the following expansion
\be\n{sU}
{\cal U}=\su U_{\sigma}^{(2n)}\, \psi_{\sigma}^{2n}\, .
\ee
Here $U_{\sigma}^{(2n)}$ are functions of angle $\theta$. The operator
$D_{\psi}$ in \eq{Dpsi} has the same form $D_{\sigma}$ for both the variables
$\psi_{\sigma}$
\be
D_{\sigma}=\partial^2_{\psi_{\sigma}}+\cot \psi_{\sigma}\,
\partial_{\psi_{\sigma}}\, .
\ee
Using the series expansion for $\cot \psi_{\sigma}$
\ba
\cot\psi_{\sigma}=\psi_{\sigma}^{-1}[1-\sum_{m=1}^{\infty} c_{2m}
\psi_{\sigma}^{2m}]\, ,\\
c_{2m}=\frac{(-1)^{m-1} 2^{2m} B_{2m}}{(2m)!}\, ,
\ea
where $B_{2m}$ are the Bernoulli numbers
\be
B_2=\frac{1}{6}\hhh B_4=-\frac{1}{30}\hhh
B_6=\frac{1}{42}\, \ldots \, ,
\ee
the relation
\be
D_{\sigma}\psi_{\sigma}^{2n}=4n^2\psi_{\sigma}^{2(n-1)}-
2n\sum_{m=1}^{\infty}c_{2m}\psi_{\sigma}^{2(n+m-1)},\, 
\ee
and equation \eq{DD} one obtains
\ba
U_{\sigma}^{(0)}&=&u_{\sigma}\, ,\\
U_{\sigma}^{(2)}&=&\frac{1}{4}(u_{\sigma,\theta\theta}+\cot\theta u_{\sigma,\theta})\, ,\\
U_{\sigma}^{(4)}&=&\frac{1}{16}\left(U_{\sigma,\theta\theta}^{(2)}+\cot\theta U_{\sigma,\theta}^{(2)}+
\frac{2}{3}U_{\sigma}^{(2)}\right)\, ,\\
U_{\sigma}^{(2n+2)}&=&\frac{1}{4(n+1)^2} [
D_{\theta}U_{\sigma}^{(2n)}\nonumber\\
&+&2\sum_{m=1}^{n} (n-m+1) c_{2m}U_{\sigma}^{(2(n-m+1))}]\,.\nonumber\\ 
\ea 

Similarly, the asymptotic expression for $\hv$ near the horizon and
singularity can be written in the form
\be\n{sV}
\hv=\sum_{n=0}^{\infty} V_{\sigma}^{(2n)} \psi_{\sigma}^{2n}\, ,
\ee
where $V_{\sigma}^{(2n)}$ are functions of the angular variable $\theta$.
Substituting expansion \eq{sU} into equation \eq{intv} one can
determine the functions $V_{\sigma}^{(2n)}$. The first three of these
functions are
\ba
V_{\sigma}^{(0)}&=&2\sigma u_{\sigma}\, ,\\
V_{\sigma}^{(2)}&=&2\sigma U_{\sigma}^{(2)}-\sigma\cot\theta
u_{\sigma,\theta}+\frac{1}{2}(u_{\sigma,\theta})^2\nonumber\\
&=&\frac{1}{2}(\sigma[u_{\sigma,\theta\theta}-\cot\theta u_{\sigma,\theta}]+(u_{\sigma,\theta})^2)\, ,\\
V_{\sigma}^{(4)}&=&\frac{1}{12}\{\sigma 
[24 U_{\sigma}^{(4)}+6(1+2\cot^2\theta) U_{\sigma}^{(2)}
\nonumber\\
&-&6\cot\theta U^{(2)}_{\sigma,\theta}
-(5+6\cot^2\theta)\cot\theta\, u_{\sigma,\theta}]\nonumber\\
&+&6u_{\sigma,\theta}U^{(2)}_{\sigma,\theta}
+(1+ 3\cot^2\theta)(u_{\sigma,\theta})^2
\nonumber\\
&-& 12\cot\theta U_{\sigma}^{(2)}u_{\sigma,\theta}+ 12(U_{\sigma}^{(2)})^2\}\, .
\ea

\section{\n{C}Geodesic motion near the singularity}

For a free particle moving in the black hole interior there exist two
integrals of motion connected with the spacetime symmetry
\be
E=-p_T=-\xi_{(T)}^{\mu}p_{\mu}\hh
L=p_{\phi}=\xi_{(\phi)}^{\mu}p_{\mu}\, .
\ee
The first has the meaning of the conserved momentum along $T$ axis,
and the second one is the angular momentum.

Consider a point $(\psi_-^0,\theta^0,T^0,\phi^0)$ near the singularity of
a distorted black hole. What is the proper time $\tau^0$ required to
fall from this point to the singularity? This time depends on
the value of $E$ and $L$. We consider the proper time for the special
value of these parameters $E=L=0$. In this case, for a moving particle
$T=const$ and $L=const$. For the Schwarzschild geometry this is the
radial motion. We also call the motion in the interior of
a disported black hole {\em `radial'} when $E=L=0$. This type of motion is a
geodesic in 2D metric
\be\n{gam}
d\gamma^2=B_-(d\theta^2-d\psi_-^2)\, ,
\ee
obtained by dimensional reduction from \eq{msing}.

Let us denote
\be
\alpha=\frac{1}{2}(\ln B_-)_{,\psi_-}\hh
\beta=\frac{1}{2}(\ln B_-)_{,\theta}\, .
\ee
Then the Christoffel symbols for the metric $d\gamma^2$ are
\ba
{}&&\Gamma^{\psi_-}_{\psi_-\psi_-}=\Gamma^{\theta}_{\theta\psi_-}
=\Gamma^{\psi_-}_{\theta\theta}=\alpha\, ,\\
{}&&\Gamma^{\theta}_{\psi_-\psi_-}=\Gamma^{\theta}_{\theta\theta}
=\Gamma^{\psi_-}_{\theta\psi_-}=\beta\, .
\ea
The geodesic equation
\be
\frac{d^2x^{\mu}}{d\tau^2}+\Gamma^{\mu}_{\nu\lambda}\frac{dx^{\nu}}{d\tau}
\frac{dx^{\lambda}}{d\tau}=0
\ee
in the metric \eq{gam} takes the form
\ba
& &\ddot{\psi}_-+\alpha(\dot{\psi}_-^2+\dot{\theta}^2)+2\beta
\dot{\psi}_-\dot{\theta}=0\, ,\n{ps}\\
& &\ddot{\theta}_-+\beta(\dot{\psi}_-^2+\dot{\theta}^2)+2\alpha
\dot{\psi}_-\dot{\theta}=0\, .\n{th}
\ea
Here the overdot denotes derivative with respect to the proper time
$\tau$. These equations obey the constraint
\be
B_-(\dot{\psi}_-^2-\dot{\theta}^2)=1\, ,\n{co}
\ee
that is the normalization condition, $u_{\mu}u^{\mu}=-1$, for $4$-velocity.

Using \eq{msing1} for the metric near the singularity we have
\be\n{lead}
\ln B_-= 4\ln\psi_- -6u_- -\ln 16+ O(\psi_-^2)\, .
\ee
Thus, in the leading order $\alpha\approx 2/\psi_-$, and $\beta\approx -3u_{-,\theta}$.

In the leading order the geodesic equations \eq{ps}-\eq{th} and the
constraint \eq{co} take the form
\ba
{}& &\psi_-\ddot{\psi}_-+2(\dot{\psi}_-^{2}+\dot{\theta}^{2})-6 u_{-,\theta} \psi_-\dot{\psi}_-\dot{\theta}
\approx0,\label{eq19aa}\\
{}& &\psi_-\ddot{\theta}-3
u_{-,\theta}\psi_-(\dot{\psi}_-^{2}+\dot{\theta}^{2})
+4\dot{\psi}_-\dot{\theta}\approx0,\label{eq19bb}\\
{}& &e^{-6u_{-}}\psi_-^{4}\left(\dot{\psi}_-^{2}
-\dot{\theta}^{2}\right)\approx16.\label{eq19cc}
\ea
According to \eq{lead}, the order of approximation in the geodesic equations corresponds to the order of approximation of the metric \eq{msing}.

We use the ambiguity in the choice of  $\tau$ to put $\tau=0$ at the
singularity for each of the `radial' geodesics approaching the
singularity. Since $\tau$ grows along geodesics directed
to the singularity, it is negative before the geodesic approaches the
singularity. The point $\tau=0$ is a singular point of the equations
\eq{eq19aa}-\eq{eq19cc}. To find approximate solution to the geodesic equations one can apply the method of asymptotic splittings described in \cite{CotBar}. A `radial' geodesic approaching the
singularity is uniquely determined by the limiting value
$\vartheta=\theta(\tau=0)$. The asymptotic expansion of $\psi_-$ and
$\theta$ near $\tau=0$ is of the form
\ba
\psi_-&=&\tta^{1/3}+\frac{2}{5} u^2_{-,\vartheta}(\vartheta)\tta+O(\tta^{4/3})\, ,\n{pt}\\
\theta&=&\vartheta+ \frac{1}{2} u_{-,\vartheta}(\vartheta)\tta^{2/3}+O(\tta^{4/3})\, \n{tt},
\ea
where $\tta=-12 e^{3u_{-}(\vartheta)} \tau$.

\end{document}